\documentclass[usenatbib]{mn2e}
\usepackage{natbib}
\usepackage{epsfig}

\catcode`\@=11
\def\gta{\ifmmode{\,\mathrel{\mathpalette\@versim>\,}}
    \else{$\,\mathrel{\mathpalette\@versim>}\,$}\fi}
\def\lta{\ifmmode{\,\mathrel{\mathpalette\@versim<\,}}
    \else{$\,\mathrel{\mathpalette\@versim<}\,$}\fi}
\def\@versim#1#2{\lower 2.9truept \vbox{\baselineskip 0pt \lineskip
    0.5truept \ialign{$\m@th#1\hfil##\hfil$\crcr#2\crcr\sim\crcr}}}
\catcode`\@=12  
\renewcommand{\[}{\begin{equation}}
\renewcommand{\]}{\end{equation}}
{\newif\ifnotend
\notendtrue
\def\veclist{ABCDEFGHIJKLMNOPQRSTUVWXYZabcdefghijklmnopqrstuvwxyz.}
\def\top#1#2.{#1}
\def\tail#1#2.{#2.}
\loop\expandafter\xdef\csname v\expandafter\top\veclist\endcsname%
{{\noexpand\bf\expandafter\top\veclist}}
\edef\veclist{\expandafter\tail\veclist}
\if\veclist.\notendfalse\fi\ifnotend\repeat}
{\newif\ifnotend
\notendtrue
\def\veclist{ABCDEFGHIJKLMNOPQRSTUVWXYZ.}
\def\top#1#2.{#1}
\def\tail#1#2.{#2.}
\loop\expandafter\xdef\csname c\expandafter\top\veclist\endcsname%
{{\noexpand\cal\expandafter\top\veclist}}
\edef\veclist{\expandafter\tail\veclist}
\if\veclist.\notendfalse\fi\ifnotend\repeat}
\def\i{{\rm i}}

\def\Teff{T_{\rm eff}}
\def\logg{\log g}
\def\feh{[{\rm Fe/H}]}

\def\ex#1{\left\langle#1\right\rangle}

\def\kms{\,{\rm km}\,{\rm s}^{-1}}
\def\mag{\,{\rm mag}}
\def\mas{\,{\rm mas}}
\def\Myr{\,{\rm Myr}}
\def\yr{\,{\rm yr}}
\def\Gyr{\,{\rm Gyr}}
\def\K{\,{\rm K}} 
\def\pc{\,{\rm pc}}
\def\kpc{\,{\rm kpc}}

\def\e{{\rm e}}
\def\d{{\rm d}}

\def\feh{\hbox{[Fe/H]}}

\def\figref#1{Fig.~\ref{#1}}
\def\pr{p}
\def\mh{\hbox{[M/H]}}
\newcommand{\beq}{\begin{equation}}
\newcommand{\eeq}{\end{equation}}

\title[New distances to RAVE stars]
{New distances to RAVE stars}

\author[J. Binney et al.]{J. Binney$^1$\thanks{E-mail:
binney@thphys.ox.ac.uk}, B. Burnett$^1$, G. Kordopatis$^2$, 
P.J. McMillan$^1$, S. Sharma$^3$, \newauthor
T. Zwitter$^4$, O. Bienaym\'e$^6$, J. Bland-Hawthorn$^3$,
M. Steinmetz$^7$, G. Gilmore$^2$,\newauthor
 M.E.K. Williams$^7$, J. Navarro$^8$, 
E.K. Grebel$^{9}$,
A. Helmi$^{10}$, 
Q. Parker$^{11}$, \newauthor
W.A. Reid$^{11}$, 
G. Seabroke$^{12}$, 
F. Watson$^{13}$,
R.F.G. Wyse$^{14}$\newauthor
\\
$^1$ Rudolf Peierls Centre for Theoretical Physics, Keble Road, Oxford OX1
3NP, UK\\ 
$^2$ Institute of Astronomy, Madingley Road, Cambridge CB3 0HA, UK\\
$^3$ Sydney Institute for Astronomy, University of Sydney, School of Physics
A28, NSW 2006, Australia\\
$^4$ University of Ljubljana, Faculty of Mathematics and Physics, Jadranska
19, 1000 Ljubljana, Slovenia and\\ Center of Excellence SPACE-SI, 
A\v{s}ker\v{c}eva cesta 12, 1000, Ljubljana, Slovenia\\
$^5$ Research School of Astronomy and Astrophysics, Australian National
University, Cotter Rd., ACT, Canberra, Australia\\
$^6$ Observatoire Astronomique de Strasbourg, 11 rue de l'Universit\'e,
Strasbourg, France\\
$^7$ Leibniz-Institut für Astrophysik Potsdam (AIP), An der Sternwarte 16,
14482 Potsdam, Germany\\ 
$^8$ Department of Physics \& Astronomy, University of Victoria, 3800
Finnerty Rd., Victoria, Canada V8P 5C2 \\
$^{9}$ Astronomisches Rechen-Institut, Zentrum f\"ur Astronomie der
Universit\"at Heidelberg, M\"onchhofstr 12-14,\\ D-69120, Heidelberg,
Germany\\
$^{10}$ Kapteyn Astronomical Institut, University of Groningen, Landleven
12, 9747 AD, Groningen, The Netherlands\\
$^{11}$ Macquarie University, Sydney, Australia\\
$^{12}$ Mullard Space Science Laboratory, University College London,
Holmbury St Mary, Dorking, RH5 6NT, UK\\
$^{13}$ Australian Astronomical Observatory, P.O. box 296, Epping, NSW 1710,
Australia\\
$^{14}$Johns Hopkins University, Departement of Physics and Astronomy, 366
Bloomberg center, 3400 N. Charles St.,\\  Baltimore, MD 21218, USA\\
}

\begin{document}

\date{Draft, September 26, 2013}

\pagerange{\pageref{firstpage}--\pageref{lastpage}} \pubyear{2012}

\maketitle

\label{firstpage}

\begin{abstract}
Probability density functions are determined from new stellar parameters for
the distance moduli of stars for which the RAdial Velocity Experiment (RAVE)
has obtained spectra with $S/N\ge10$.  Single-Gaussian fits to the pdf in
distance modulus suffice for roughly half the stars, with most of the other
half having satisfactory two-Gaussian representations. As expected,
early-type stars rarely require more than one Gaussian.  The expectation
value of distance is larger than the distance implied by the expectation of
distance modulus; the latter is itself larger than the distance implied by
the expectation value of the parallax. Our parallaxes of Hipparcos stars
agree well with the values measured by Hipparcos, so the expectation of
parallax is the most reliable distance indicator.  The latter are improved by
taking extinction into account.  The effective temperature absolute-magnitude
diagram of our stars is significantly improved when these pdfs are used to
make the diagram. We use the method of kinematic corrections devised by
Sch\"onrich, Binney \& Asplund to check for systematic errors for general
stars and confirm that the most reliable distance indicator is the
expectation of parallax.  For cool dwarfs and low-gravity giants
$\ex{\varpi}$ tends to be larger than the true distance by up to 30 percent.
The most satisfactory distances are for dwarfs hotter than $5500\K$. We
compare our distances to stars in 13 open clusters with cluster distances
from the literature and find excellent agreement for the dwarfs and
indications that we are over-estimating distances to giants, especially in
young clusters.
\end{abstract}

\section{Introduction}

Surveys of the stellar content of our Galaxy are key to the elucidation of
the Galaxy's structure and history. Consequently, over the last decade
considerable observational resources have been devoted to such surveys. Three
surveys are particularly worthy of note: the 2MASS survey \citep{2MASS}, the
Sloan Digital Sky Survey (SDSS) \citep{SDSS,SEGUE} and the RAdial Velocity
Experiment (RAVE) \citep{RAVE,DR3}. The 2MASS survey was an all-sky, near
infrared photometric survey, while the SDSS survey combined a
photometric survey in the $ugriz$ system with spectroscopy for a subset of
objects with spectral resolution $R=2500$. The RAVE survey has taken spectra
at resolution $R\simeq7500$ of $\sim500\,000$ stars that have 2MASS
photometry. The RAVE and SDSS surveys are complementary in that SDSS worked
at apparent magnitudes $r\gta18$ so faint that it catalogued mainly dwarf
stars that lie more than $500\pc$ from the Sun, while RAVE operates at
apparent magnitudes $I\approx9-13$ and observes both nearby dwarfs and giants
at distances up to $\sim4\kpc$ \citep{Burnettetal}.

Although the ideal way to extract science from a survey is to project models
into the space of observables, i.e., sky coordinates, line-of-sight velocity,
apparent magnitudes, etc., and fit the projected models to the data
\citep[e.g.][]{BinneyBangalore}, in practice one generally assigns a distance
to each star and uses this distance to place the star in the space in which
physics applies, namely phase space complemented with luminosity, colour,
chemical composition, etc. Since RAVE's targets overwhelmingly lie beyond the
range of Hipparcos and include both dwarfs and giants, the task of assigning
distances to these stars is complex. To date three papers
\citep{Breddels,Zwitter10,Burnettetal} address this task with techniques of
increasing sophistication. Results presented in those papers are based on
stellar parameters produced by the pipeline that was developed for analysis
of the RAVE spectra. This pipeline was described in the papers that
accompanied the second and third releases of RAVE data \citep{DR2,DR3}.
Between those two data releases changes were made to the pipeline's
parameters that were designed to improve the accuracy of the derived
metallicities, but the parameters from neither version of the pipeline were
entirely satisfactory \citep[][hereafter B11]{Burnettetal}.

On account of residual internal and external inconsistencies in the
parameters, a completely new pipeline has been developed for the analysis of RAVE
spectra.  This pipeline and the stellar parameters it produces are described
in \cite{DR4}. The new stellar parameters form a much more compelling
and consistent database than the old ones, and their arrival prompts us to
revisit the assignment of distances using the new parameters as inputs.

We use the Bayesian framework described by \cite{BurnettB} but modified to allow
for the impact of interstellar dust. Two other significant novelties are (i)
the production of multi-Gaussian fits to each star's probability density function
(pdf) in distance modulus and (ii) the use of the kinematic correction
factors introduced by \cite{SBA} to check for systematic errors in our
distances. We have derived distances for all stars that have spectra to
which the new pipeline assigns a signal-to-noise ratio of 10 or higher. When
a star has more than one spectrum in the database, the catalogued distance is
that derived from the highest S/N spectrum.

The plan of the paper is as follows. In Section~\ref{sec:method} we
recapitulate the principles of Bayesian distance determination and describe
how we take extinction into account. In Section~\ref{sec:pdfs} we discuss
typical pdfs in distance modulus and explain how we produce multi-Gaussian
fits to them. In Section~\ref{sec:Hipp} we compare our spectrophotometric
parallaxes to Hipparcos parallaxes and ask how these comparisons are affected
by neglecting extinction. In Section~\ref{sec:all} we analyse our distances
to the generality of stars, using kinematic correction factors
to test for systematic biases in distances as functions of surface gravity
or effective temperature, and to modify distance pdfs
(Section~\ref{sec:SBA}). In Section~\ref{sec:clusters} we compare our
distances to cluster stars with the established distances to their clusters.
In Section \ref{sec:repeat} we examine the scatter in the distances to the
same star obtained from different spectra.  In Section \ref{sec:Av} we
examine the distribution of extinctions to stars. Section~\ref{sec:discuss}
sums up.

\section{Methodology}\label{sec:method}

As in B11 we start from the trivial Bayesian statement
\[
\pr(\hbox{model}|\hbox{data})={\pr(\hbox{data}|\hbox{model})\pr(\hbox{model})
\over\pr(\hbox{data})},
\]
 where ``data'' comprises the observed parameters and photometry of an
individual star and ``model'' comprises a star of specified initial mass
$\cM$, age $\tau$, metallicity $\hbox{[M/H]}$, and location.  We use $p({\rm
model|data})$ either to calculate expectation values $\ex{x}$ and dispersions
$\sigma_x$ of quantities of interest, such as the stars's distance $x=s$ and
parallax $x=\varpi$, by integrating $P({\rm model|data})$ times an
appropriate power of $x$ through the space spanned by the model parameters
$\hbox{[M/H]},\tau,\cM,\ldots$, or the pdf in distance
modulus by marginalising $P({\rm model|data})$ over all model parameters
other than distance.

A key role is played by the prior probability $\pr(\hbox{model})$, which
reflects our prior knowledge of the Galaxy: massive young stars are rarely
found far from the plane, while a star far from the plane is likely to be old
and have sub-solar abundances.  We have used the same three-component prior
used in B11:
 \begin{equation}\label{eq:priorofx}
  p(\hbox{model}) = p(\cM) \sum_{i=1}^3 p_i(\mh) \, p_i(\tau) \, p_i(\mathbf{r}),
\end{equation}
 where $i=1,2,3$ correspond to a thin disc, thick disc and stellar halo,
respectively. We assumed an identical Kroupa-type IMF for all three
components and distinguish them as follows:

\paragraph*{Thin disc ($i=1$):}
\begin{eqnarray} \label{eq:thindisc}
  p_1(\mh) &=& G(\mh, 0.2), \nonumber \\
  p_1(\tau)  &\propto& \exp(0.119 \,\tau/\mbox{Gyr}) \quad \mbox{for $\tau \le 10$\,Gyr,}  \\
  p_1(\mathbf{r}) &\propto& \exp\left(-\frac{R}{R_d^{\rm{thin}}} - \frac{|z|}{z_d^{\rm{thin}}}  \right);  \nonumber
\end{eqnarray}

\paragraph*{Thick disc ($i=2$):}
\begin{eqnarray}\label{eq:thickdisc}
  p_2(\mh) &=& G(\mh+0.6, 0.5), \nonumber \\
  p_2(\tau)  &\propto& \mbox{uniform in range $8 \le \tau \le 12$\,Gyr,} \\
  p_2(\mathbf{r}) &\propto& \exp\left(-\frac{R}{R_d^{\rm{thick}}} - \frac{|z|}{z_d^{\rm{thick}}}  \right); \nonumber
\end{eqnarray}

\paragraph*{Halo ($i=3$):}
\begin{eqnarray}
  p_3(\mh) &=& G(\mh+1.6, 0.5), \nonumber \\
  p_3(\tau)  &\propto& \mbox{uniform in range $10 \le \tau \le 13.7$\,Gyr,} \\
  p_3(\mathbf{r}) &\propto& r^{-3.39}; \nonumber
\end{eqnarray}
 where $R$ signifies Galactocentric cylindrical radius, $z$ cylindrical
height and $r$ spherical radius, and $G(x,y)$ is a Gaussian distribution in
$x$ of zero mean and dispersion $y$. The parameter values were taken as in
Table~\ref{table:params}; the values are taken from the analysis of SDSS data
in \cite{Juric_cut}. The metallicity and age distributions for the thin disc
come from \cite{Haywood} and \cite{Aumer}, while the radial density of the
halo comes from the `inner halo' detected in \cite{Carollo}. The metallicity
and age distributions of the thick disc and halo are influenced by
\cite{Reddy} and \cite{Carollo}.

The normalizations were then adjusted so that at the solar position, taken as
$R_0=$~8.33\,kpc (\citealt{Gillessen}), $z_0=$~15\,pc
\citep{BinneyGS,Juric_cut}, we have number
density ratios $n_2 /n_1 = 0.15$ (\citealt{Juric_cut}), $n_3 /n_1 = 0.005$
(\citealt{Carollo}).

\begin{table}
  \begin{center}
    \caption{Values of disc parameters used.\label{table:params}}
    \begin{tabular}{cr}
      \hline
      Parameter & Value (pc) \\
      \hline
      $R_d^{\rm{thin}}$  & 2\,600 \\[3pt]
      $z_d^{\rm{thin}}$  & 300 \\[3pt]
      $R_d^{\rm{thick}}$ & 3\,600 \\[3pt]
      $z_d^{\rm{thick}}$ & 900 \\
      \hline
    \end{tabular}
  \end{center}
\end{table}

\begin{table}
  \begin{center}
    \caption{Metallicities of isochrones used, taking $(Z_\odot , Y_\odot) = (0.017,0.260)$.\label{table:Zs}}
    \begin{tabular}{rrr}
      \hline
      $Z$ & $Y$ & $\mh$ \\
      \hline
      0.0022 & 0.230 & $-0.914$ \\
      0.003 & 0.231 & $-0.778$ \\
      0.004 & 0.233 & $-0.652$ \\
      0.006 & 0.238 & $-0.472$ \\
      0.008 & 0.242 & $-0.343$ \\
      0.010 & 0.246 & $-0.243$ \\
      0.012 & 0.250 & $-0.160$ \\
      0.014 & 0.254 & $-0.090$ \\
      0.017 & 0.260 & 0.000 \\
      0.020 & 0.267 & 0.077 \\
      0.026 & 0.280 & 0.202 \\
      0.036 & 0.301 & 0.363 \\
      0.040 & 0.309 & 0.417 \\
      0.045 & 0.320 & 0.479 \\
      0.050 & 0.330 & 0.535 \\
      0.070 & 0.372 & 0.727 \\
      \hline
    \end{tabular}
  \end{center}
\end{table}

The IMF chosen follows the form originally proposed by \cite{Kroupa}, with a
minor modification following \cite{Aumer}, being
 \begin{equation}
  p(\cM) \propto \cases{\cM^{-1.3}&if $\cM<0.5\,$M$_\odot$,\cr
    0.536 \, \cM^{-2.2}&if $0.5\,$M$_\odot \le \cM<1\, $M$_\odot$,\cr
    0.536 \, \cM^{-2.519}&otherwise.}
\end{equation}

We predicted the photometry of stars from  the isochrones of the Padova group
(\citealt{Bertelli}), which provide tabulated values for the observables of
stars with metallicities ranging upwards from around $\mh \approx -0.92$,
ages in the range $\tau \in [0.01, 19]$\,Gyr and masses in the range $\cM \in
[0.15, 20]$\,M$_\odot$. We used isochrones for 16 metallicities as shown in
Table~\ref{table:Zs}, selecting the helium mass fraction $Y$ as a function of
metal mass fraction $Z$ according to the relation used in \cite{Aumer}, i.e.\
$Y \approx 0.225 + 2.1 Z$ and assuming solar values of $(Y_\odot , Z_\odot) =
(0.260,0.017)$. The metallicity values were selected by eye to ensure that there 
was not a great change in the stellar observables between adjacent isochrone sets.

In B11 no correction was made for the differences between the
Johnson-Cousins-Glass photometric system used for the Padova stellar models
that we use and the 2MASS system. Here we use the transformations of
\cite{Koen07} to transform the 2MASS magnitudes $J_2,\ldots$ to the
Johnston-Cousins-Glass magnitudes $J,\ldots$:
\begin{eqnarray}
J&=&0.029+J_2+0.07(J_2-K_2)-0.045(J_2-H_2)^2\nonumber\\
H&=&H_2+0.555(H_2-K_2)^2-0.441(H_2-K_2)\nonumber\\
&&\qquad+0.089(J_2-H_2)\\
K&=&0.009+K_2+0.195(J_2-H_2)^2-0.156(J_2-H_2)\nonumber\\
&&\qquad+0.304(H_2-K_2)-0.615(H_2-K_2)^2.\nonumber
\end{eqnarray}
Unless explicitly stated to the contrary, we will state $JHK$ magnitudes in the
Johnston-Cousins-Glass  system.

Dust both dims and reddens stars. Let the column of dust between us and a
given star produce optical extinction $A_V$, then from \cite{RiekeL85} we
take the extinctions to be
 \begin{eqnarray}
A_J&=&0.282A_V\nonumber\\
A_H&=&0.175A_V\\
A_K&=&0.112A_V.\nonumber
\end{eqnarray}

In B11 $A_V$ was set identically to zero and the $H$ magnitude was not
employed. Here we include the $H$ magnitude in the set of observations so we
have three constraints on the star's spectral distribution: the
spectroscopically derived $\Teff$ and two IR colours. Consequently, we should
be able to constrain the extinction to some extent. We integrate over all
possible values of $A_V$.  We include $A_V$ in the prior by multiplying the
prior (\ref{eq:priorofx}) by the probability density of $A_V$. Since $A_V$ is
an intrinsically non-negative quantity, a completely flat prior would be one
uniform in $a\equiv\ln(A_V)$. We do not want a flat prior but one that
reflects increasing extinction with distance and higher extinction towards
the Galactic centre than towards the poles. Let $a_{\rm prior}(\vx)$ be the
expected value of $\ln(A_V)$ for the location $\vx$. Then a natural choice
for the probability of extinctions associated with the interval $(a,a+\d a)$
is
 \begin{eqnarray}
\d P&=&(2\pi\sigma^2)^{-1/2}\e^{-(a-a_{\rm prior})^2/2\sigma^2}\,\d a\nonumber\\
&=&(2\pi\sigma^2)^{-1/2}\e^{-\ln^2(A_V/A_{V{\rm prior}})/2\sigma^2}\,\d a.
\end{eqnarray}
 The dispersion $\sigma$ reflects the random fluctuation of the extinction
from one sight-line to the next on account of the cloudy nature of the
interstellar medium. We have rather arbitrarily set $2\sigma^2=1$. 

 $A_{V{\rm prior}}$ is related to distance by
 \[
A_{V{\rm prior}}(b,\ell,s)=A_{V\infty}(b,\ell){\int_0^s\d
s'\,\rho[\vx(s')] \over \int_0^\infty\d s'\,\rho[\vx(s')]}
\]
 where $\vx(s)$ is the position-vector of the point that lies distance $s$
down the line of sight $(b,\ell)$, $A_{V\infty}(b,\ell)$ is defined below
and $\rho(\vx)$ is a model of the density of extincting material. Following
\cite{Sharma11} we adopt
 \[\label{eq:dmodel}
\rho(\vx)=\exp\left[{R_0-R\over h_R}-{|z-z_{\rm w}|\over
k_{\rm fl}h_z}\right],
\]
 where $k_{\rm fl}(R)$ and $z_{\rm w}(R)$ describe the flaring and warping of the
gas disc:
\begin{eqnarray}
k_{\rm fl}(R)&=&1+\gamma_{\rm fl}\min(R_{\rm fl},R-R_{\rm fl})\nonumber\\
z_{\rm w}(R,\phi)&=&\gamma_{\rm w}\min(R_{\rm w},R-R_{\rm w})\sin\phi.
\end{eqnarray}
 Here $\phi$ is the Galactocentric azimuth that increases in the direction of
Galactic rotation and places the Sun at $\phi=0$. Table~\ref{tab:dust} gives the
values of the parameters that occur in these formulae. 

We take the extinction to infinity, $A_{V\infty}(b,\ell)$, from observation:
except along exceptionally obscured lines of sight, $A_{V\infty}$ is 3.1
times the reddening estimated by \cite{Schlegel}. However, \cite{ArceGoodman}
pointed out that the Schlegel et al.\ over-estimate the reddening in regions
with $E(B-V)>0.15$. Following \cite{Sharma11} we correct for this effect by multiplying the Schlegel et
al.\ values of $E~(B-V)$ by the correction factor
 \[
f(E(B-V))=0.6+0.2\left[1-\tanh\left({E(B-V)-0.15\over0.3}\right)\right],
\]
 which has the effect of leaving $E(B-V)$ invariant for $E(B-V)\lta0.16$ and
multiplying large values of $E(B-V)$ by a factor 0.6.

The function $A_{V{\rm prior}}(s)$ is tabulated on a
non-uniform grid in $s$ before each star is analysed so $A_{V{\rm prior}}$
can be subsequently obtained quickly by linear interpolation.

Given a model star characterised by ([M/H],$\tau,\cM$), a first estimate of
the distance to the star is made under the assumption $A_V=0$. Then $A_{V{\rm
prior}}$ is evaluated for this distance and a second estimate of distance
obtained, and $A_{V{\rm prior}}$ is evaluated at this improved distance and
stored as $A_{V{\rm model}}$. The reddened $J-K$ colour of the star is now
predicted and compared with the observed colour.  The given model star is
considered sufficiently plausible to be worth considering further only if
both its colour reddened by e times $A_{V{\rm model}}$ is redder than the
blue end of the $3\sigma$ range around the measured colour and the star's
colour reddened by $1/\e$ times $A_{V{\rm model}}$ is bluer that the red end
of the measured $3\sigma$ range. If these conditions are satisfied, we
consider values of $A_V$ that lie the range $(\e^{-1.5},\e^{1.5})A_{V{\rm
model}}$. For each value of $A_V$ all plausible distances are considered.

We calculate the expectation $\ex{a}$ of $a\equiv\ln A_V$ and use $\widetilde
A_V\equiv\exp(\ex{a})$ as our final estimate of the extinction to each star.

\begin{table}
\caption{Parameters of the model of the dust distribution. Distances are in
kiloparsecs.}\label{tab:dust}
\begin{center}
\begin{tabular}{ccccccc}
\hline
$A_V(0)$&$h_R$&$h_z$&$R_{\rm fl}$&$\gamma_{\rm fl}$&$R_{\rm w}$&$\gamma_{\rm w}$\\
1.67 & 4.2 &0.088& $1.12R_0$  & 0.0054      & 8.4 &  0.18\\
\hline
\end{tabular}
\end{center}
\end{table}

\begin{figure*}
\centerline{
\includegraphics[width=.3\hsize]{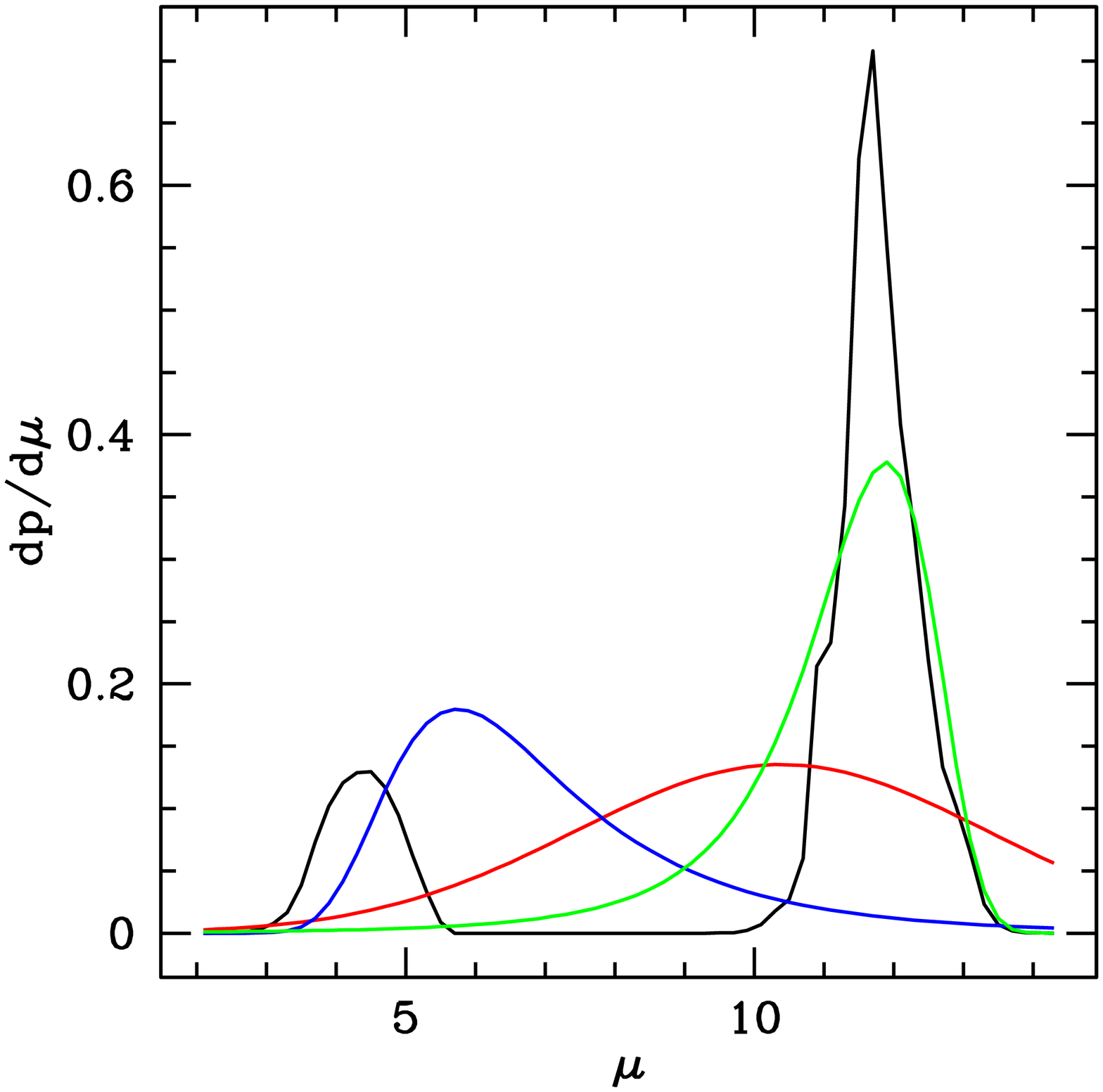}
\includegraphics[width=.3\hsize]{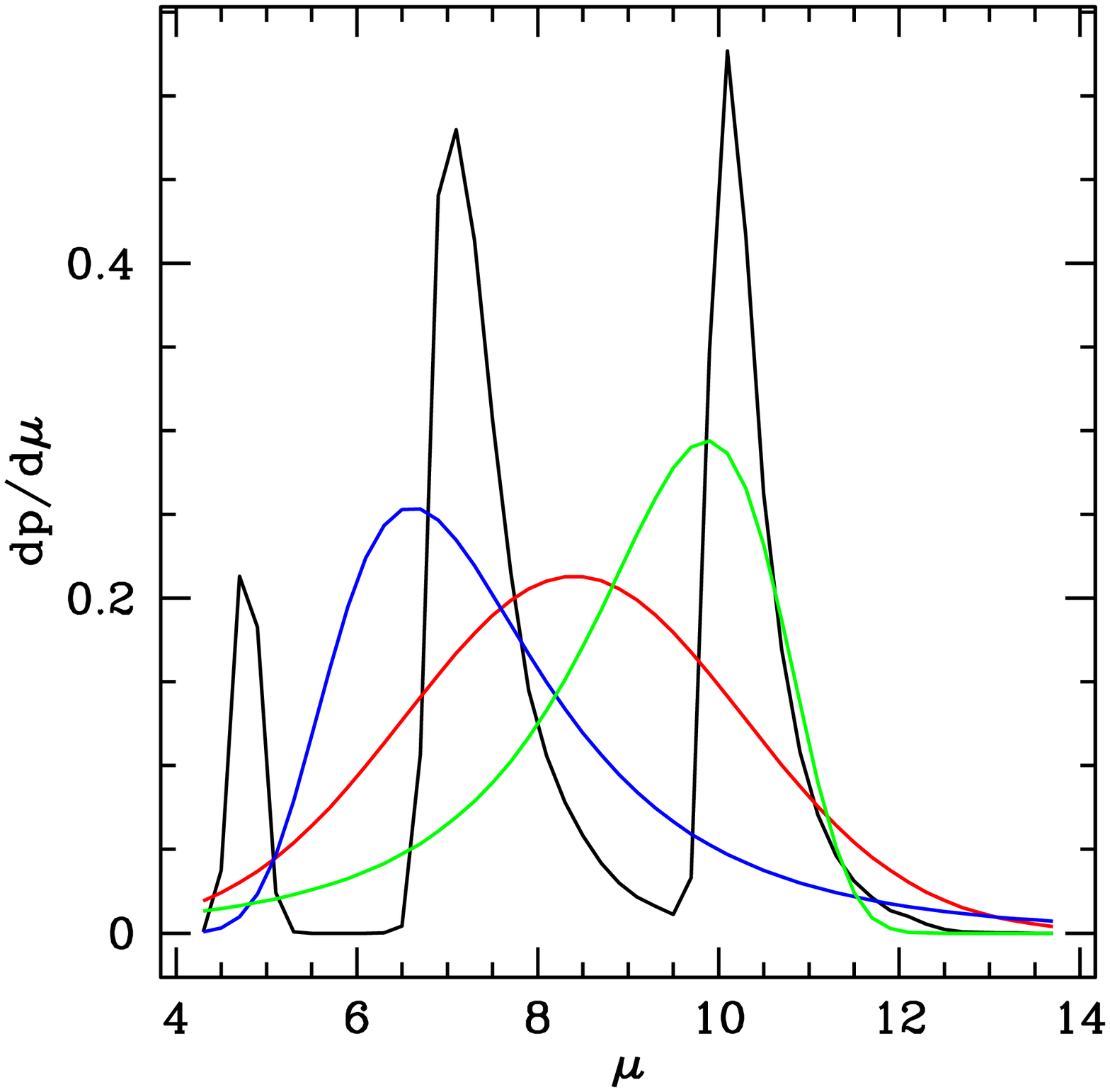}
\includegraphics[width=.3\hsize]{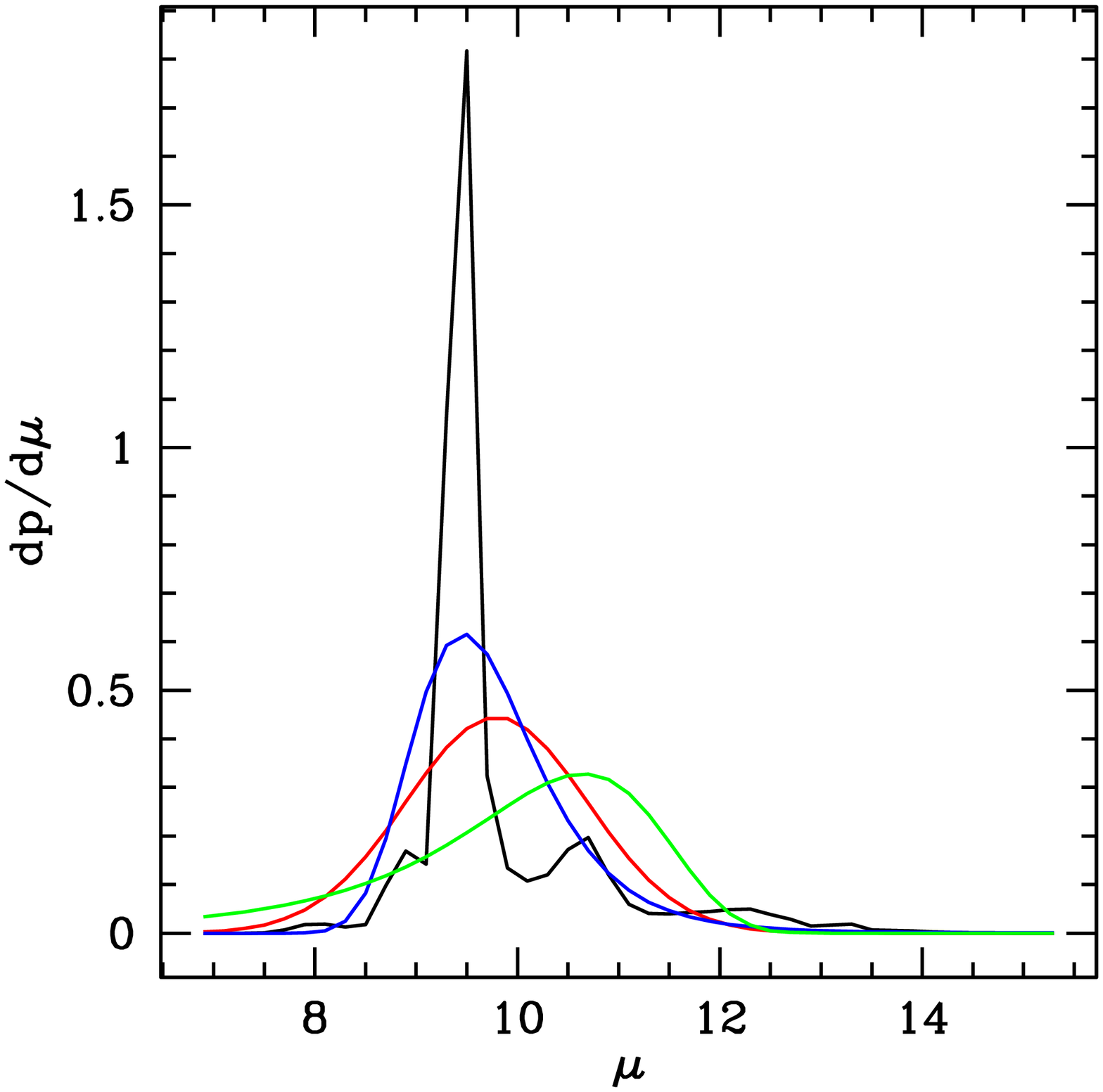}
}
 \caption{Pdfs in distance modulus for three RAVE stars that are not amoung
 the $\sim45$ per cent of stars with pdfs that can be adequately fitted by a
 single Gaussian.  The {black}
line shows the computed pdf while the red curve is a Gaussian with the same
mean and standard deviation. The blue curve is the pdf implied by a Gaussian
in parallax and the green curve is that implied by a Gaussian in distance.
Approximately 20 per cent of the pdfs are bimodal (left), 5 per cent are
trimodal (centre) and 25 per cent are dominated by a sharp peak that sits on
a broader component that has a much lower probability \emph{density}, but
contributes a significant fraction of the total probability (right).}
\label{fig:badpdfs}
\end{figure*}
\begin{figure*}
\centerline{
\includegraphics[width=.3\hsize]{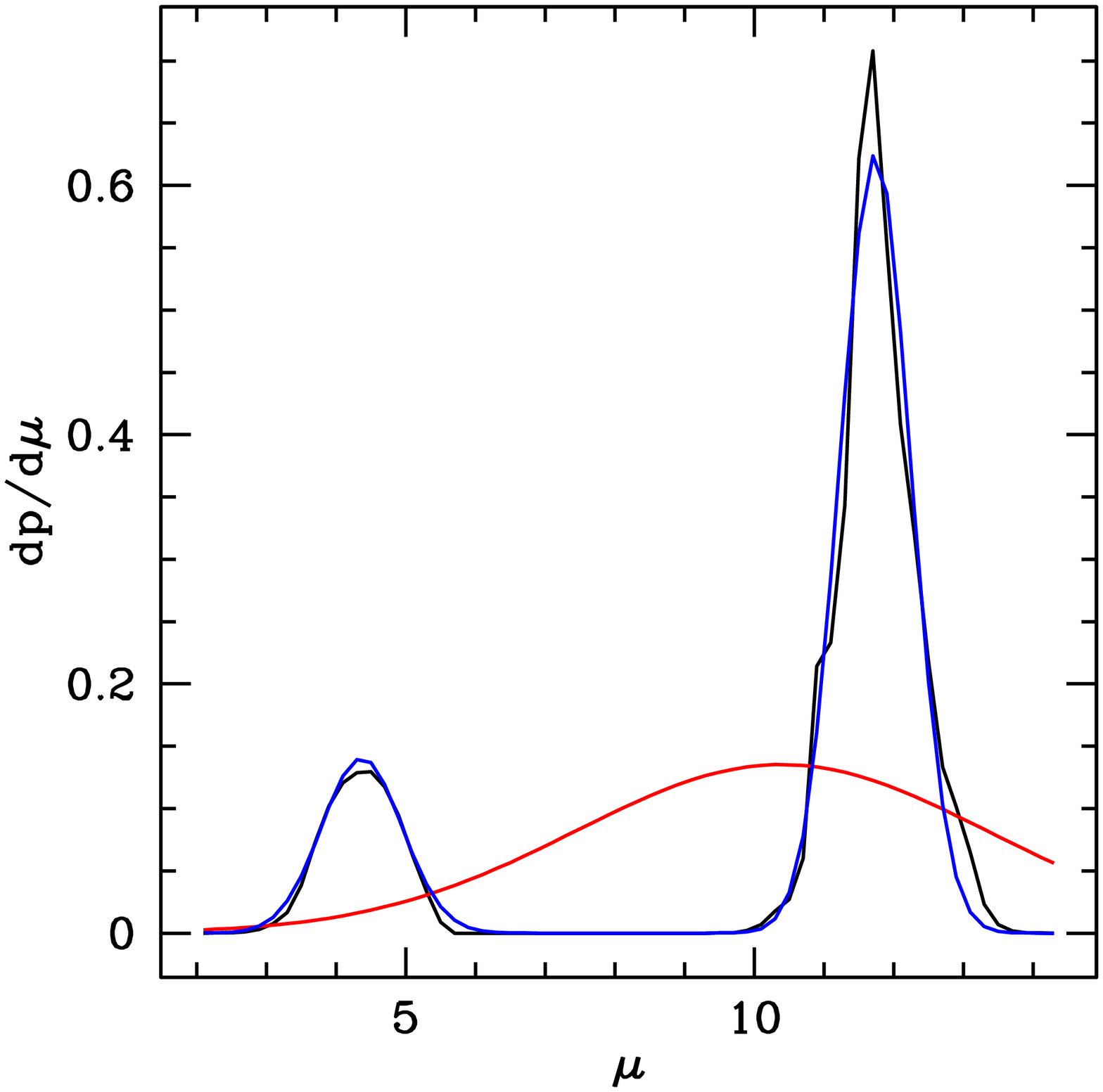}
\includegraphics[width=.3\hsize]{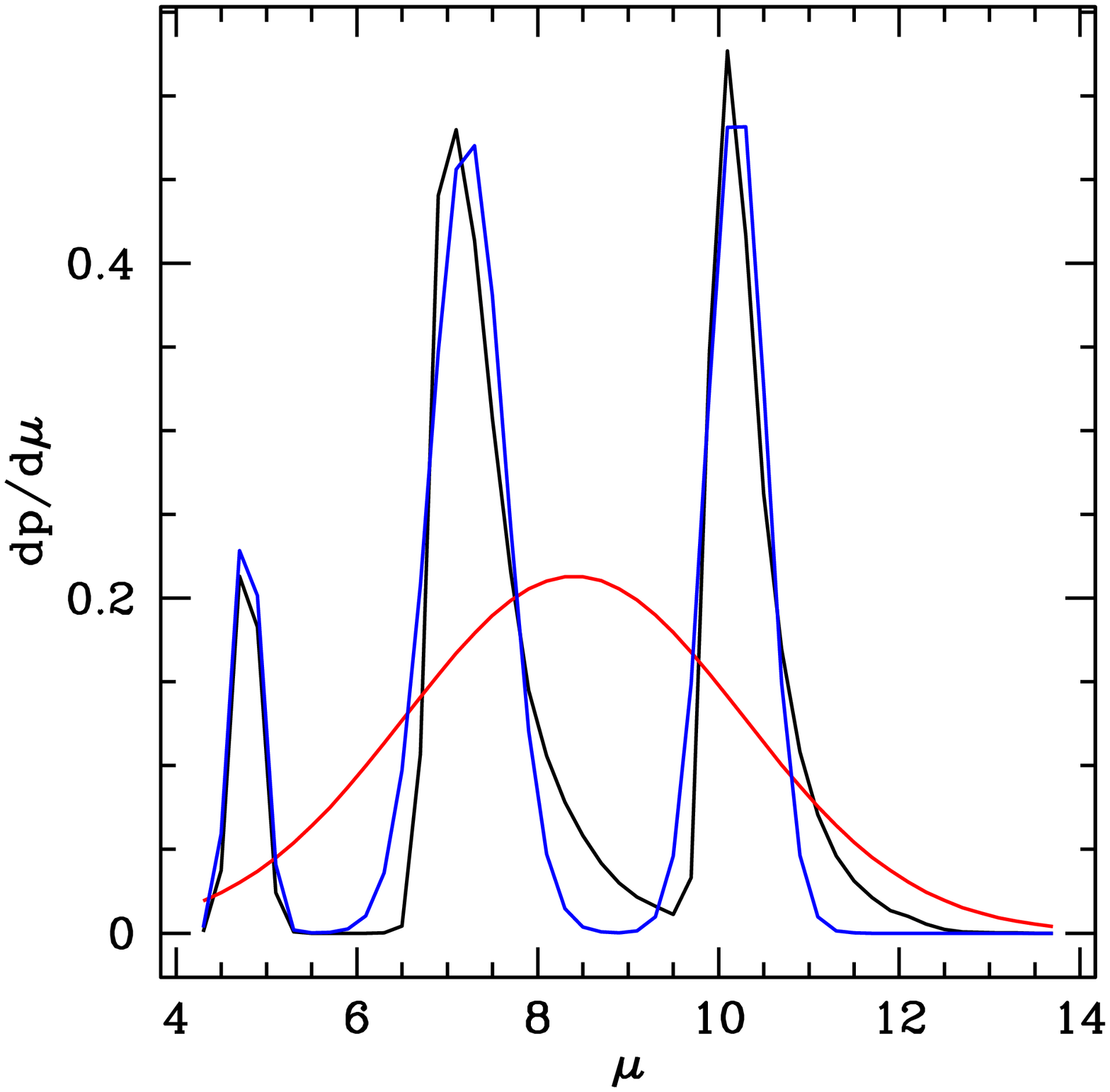}
\includegraphics[width=.3\hsize]{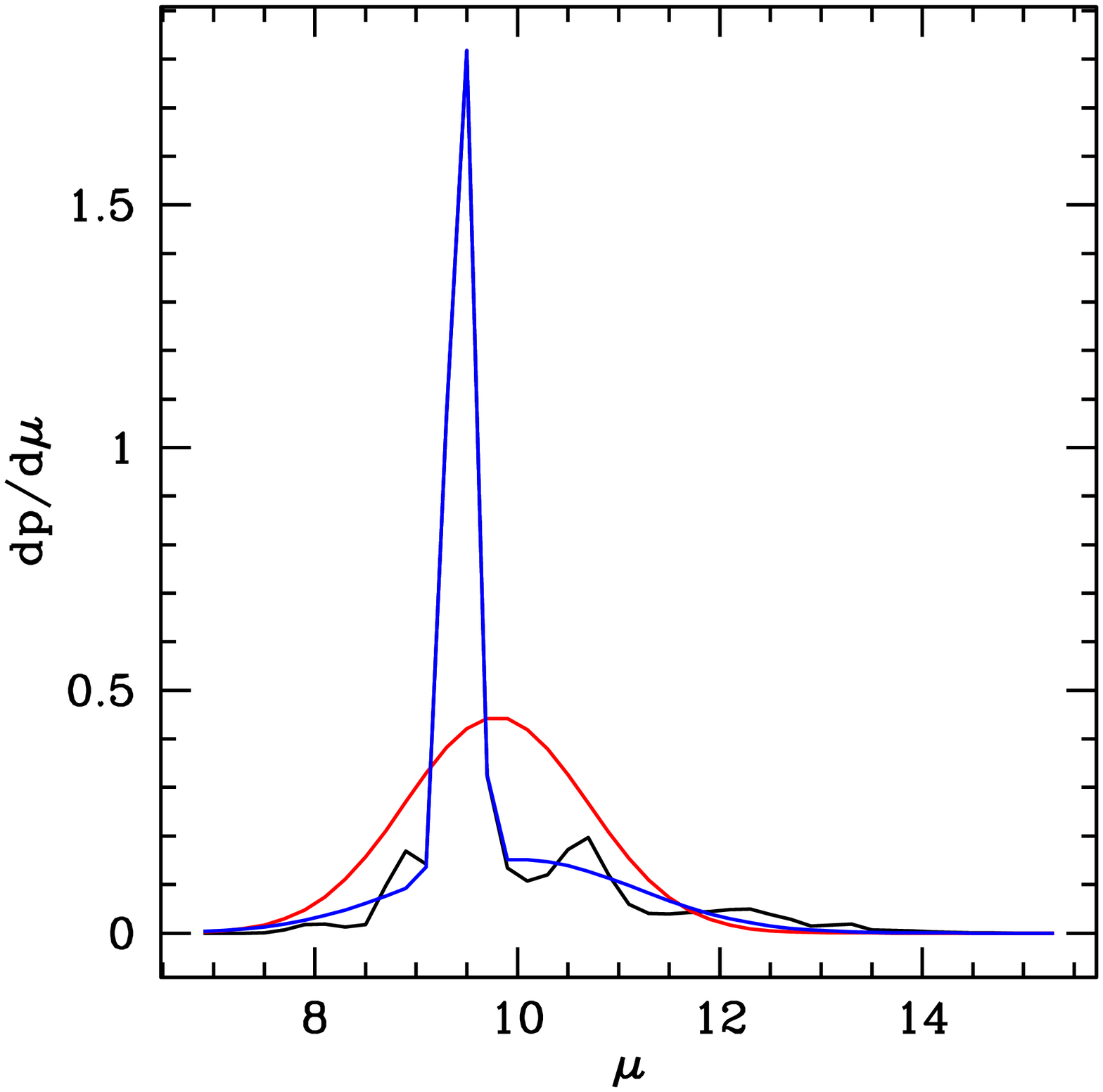}
}
 \caption{The black curves show the pdfs in distance modulus for the same
three stars as in \figref{fig:badpdfs} while 
the {blue} line shows  the chosen  multi-Gaussian fit fitting.
} \label{fig:goodpdfs}
\end{figure*}
\begin{figure*}
\centerline{
\includegraphics[width=.3\hsize]{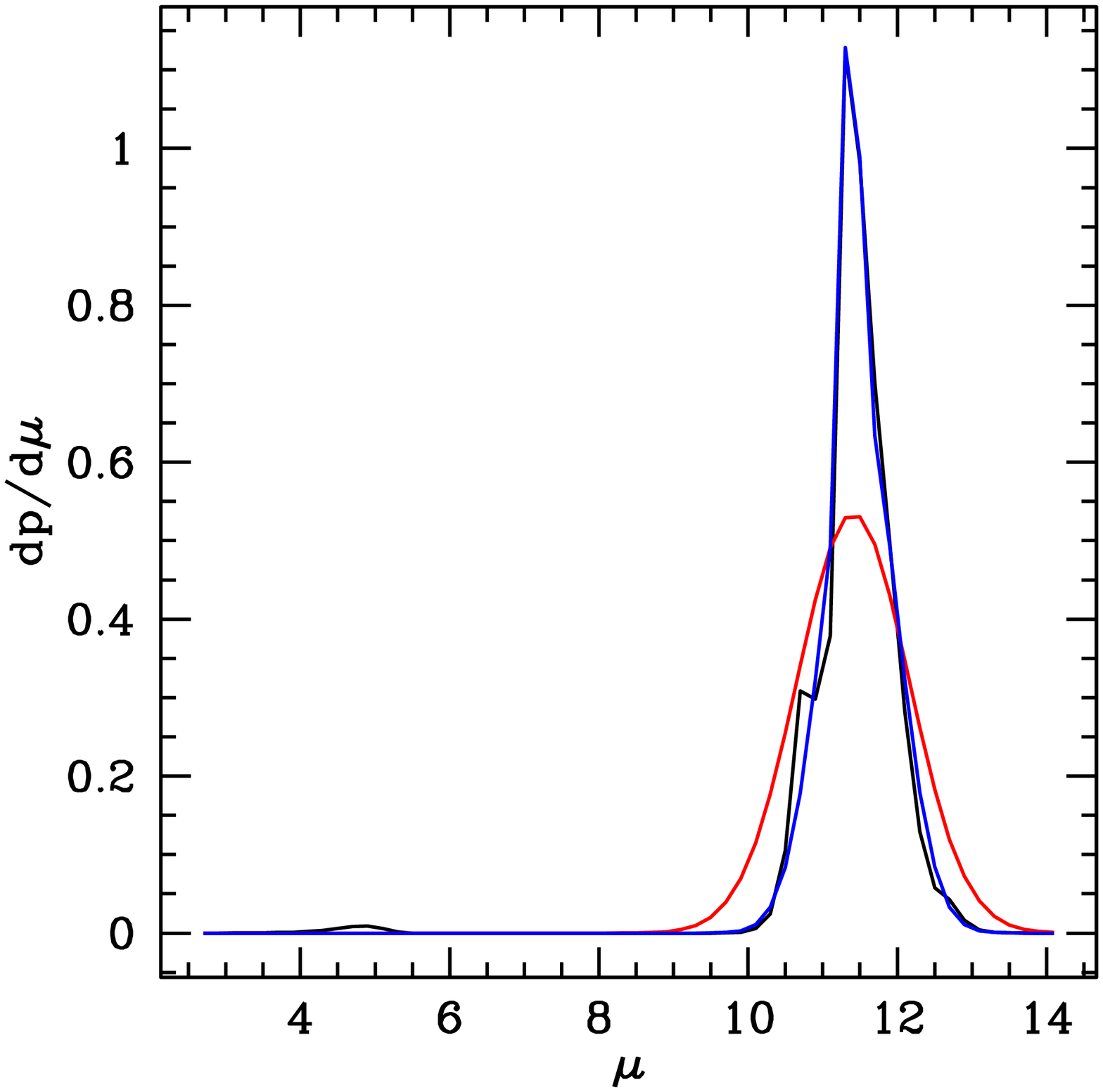}
\includegraphics[width=.3\hsize]{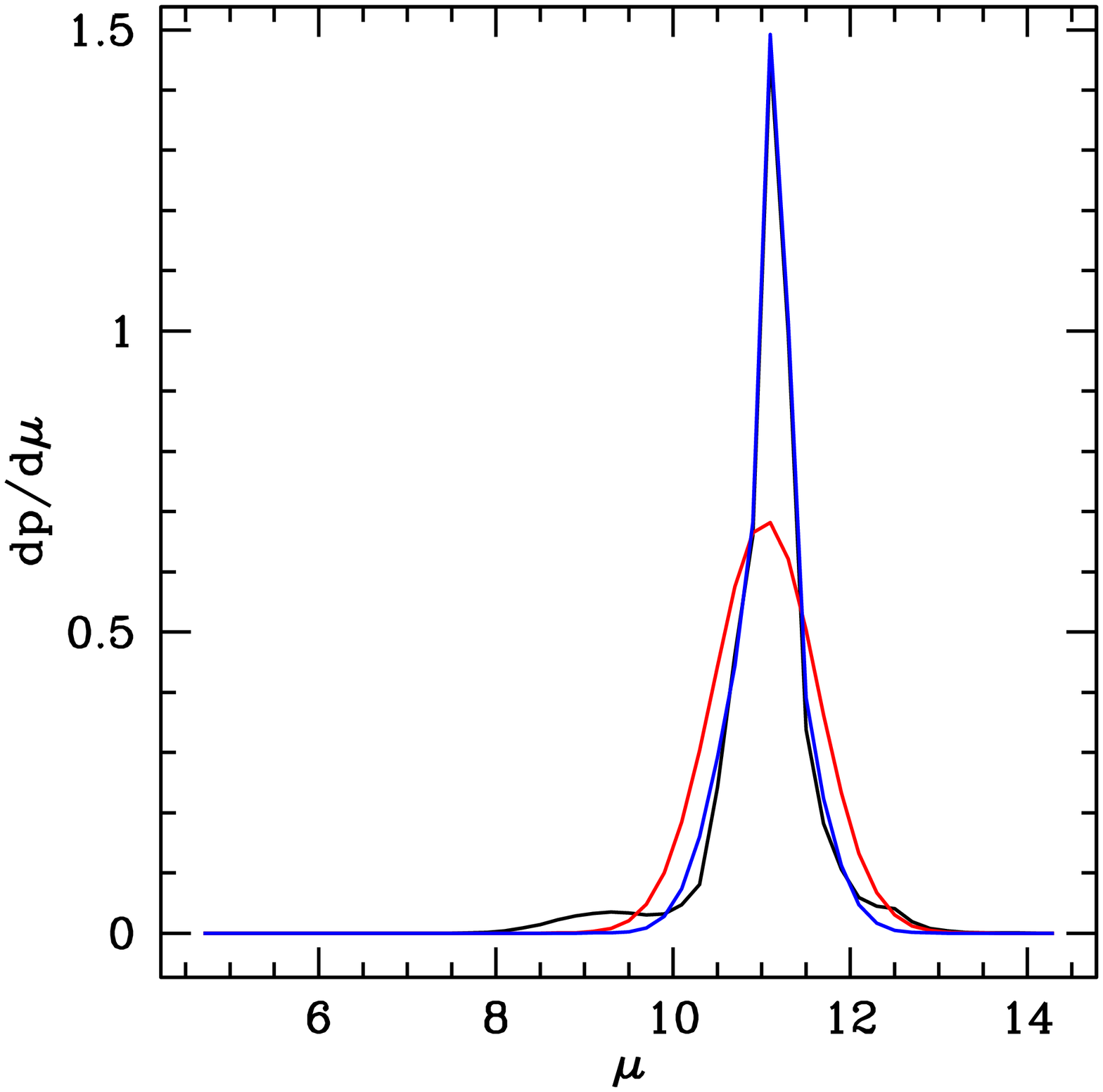}
\includegraphics[width=.3\hsize]{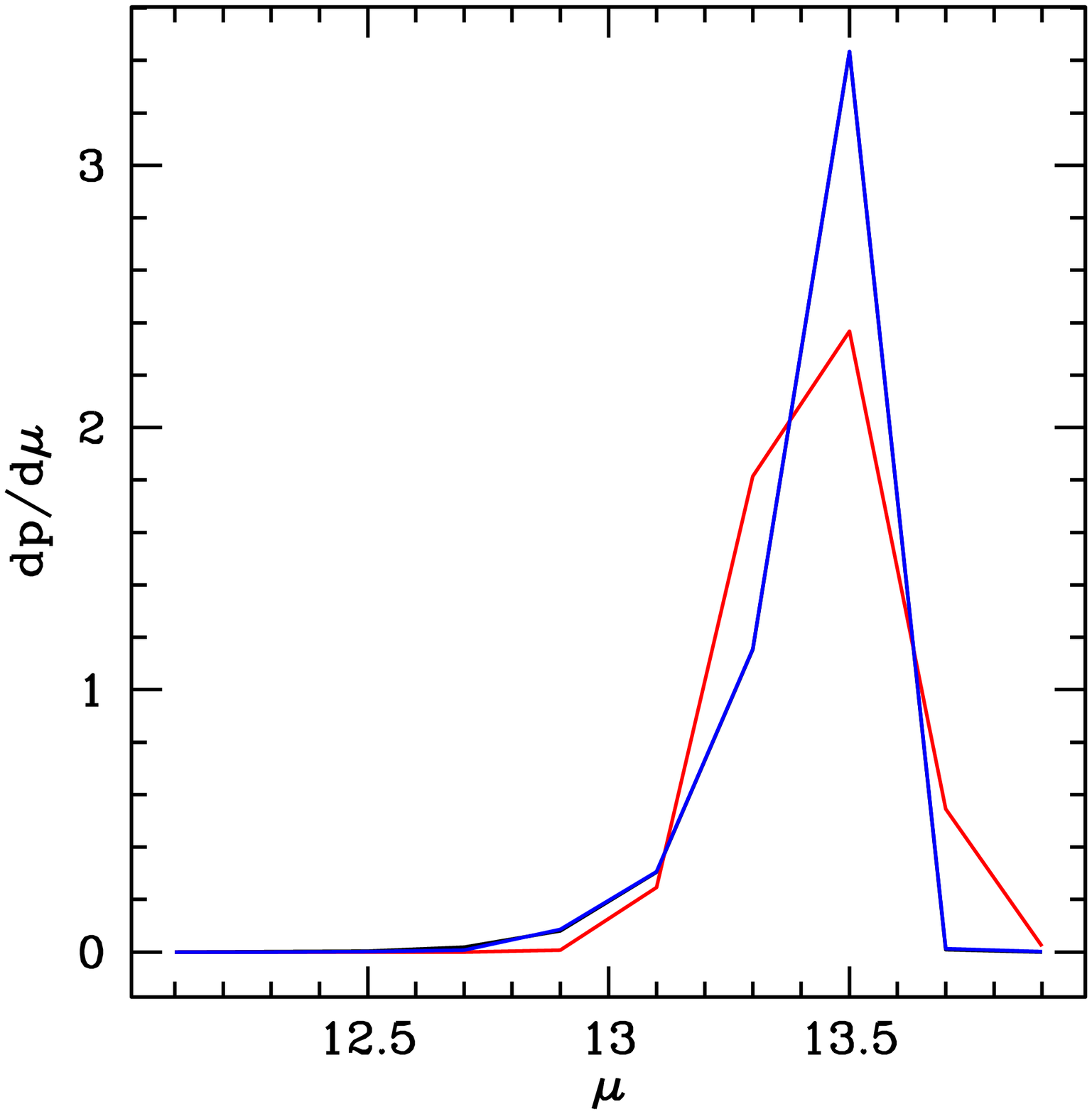}
}
 \caption{Pdfs in distance modulus for three stars that are flagged as having
potentially inadequate multi-Gaussian fits to their pdfs because the true
dispersions in $\mu$ differ by more than 20 per cent from the dispersions of
the fitted models. As in \figref{fig:goodpdfs} the {black} line
shows the computed pdf and the {blue} line shows the output from the
multi-Gaussian fitting. The {red} line shows a Gaussian with the same
mean and dispersion. The reason for the flag in the left panel is the very
small component of the computed pdf at $\mu\sim5$, which is not picked up by the
fitted pdf. In the centre panel the flag is raised because the fitted pdf
does not fully reflect the broad wings of the pdf. In the right panel the fit
appears nearly perfect, but this reflects the resolution of the histogram
(used both to find the approximation and to plot this figure) being
low compared to the width of the pdf -- the dispersion in distance modulus is
$0.15$, which is smaller than the histogram bin size.  } \label{fig:badflag}
\end{figure*}

\section{PDFs for distance}\label{sec:pdfs}

The Bayesian argument yields the five-dimensional probability density function
(pdf) that each star has a given mass, metallicity, age, line-of-sight
extinction and distance, but \cite{BurnettB} and \cite{Burnettetal} reported
only the implied means and standard deviations of distance and parallax.
Hence they had two logically independent measures of the distance to a star:
$\ex{s}$ and $1/\ex{\varpi}$. A third natural distance measure is provided by
the expectation of the distance modulus $\mu=5\log_{10}(s/10\pc)$. We shall
show that these three measures yield systematically different  distances and
conclude that $1/\ex{\varpi}$ is the most reliable estimate.

A logical next step is to inspect the pdfs we obtain for $s$, etc., after
marginalising over the star's other properties.  If any of these pdfs is well
approximated by a Gaussian, it can be fully characterised by its mean and
dispersion. In this section we show that the pdfs often deviate significantly
from a Gaussian, and in this case it is important to know more than the pdf's
mean and dispersion.

Fig.~\ref{fig:badpdfs} shows pdfs in distance modulus for
three stars. The red curves show Gaussian distributions in distance modulus
$\mu\equiv m-M$, while the green curves show distributions that are Gaussian
in distance $s$ and the blue curves show distributions that are Gaussian in
parallax $\varpi$.  Given how strongly these three curves differ from one
another, especially in the left and centre panels, it is clear that a very
particular assumption is being made if one supposes that a star's
distribution of either $\mu$, $s$ or $\varpi$ is Gaussian, and if one of
these distributions \emph{is} Gaussian, the other two cannot be.

In each panel of \figref{fig:badpdfs} the black curve shows the computed
marginalised pdf in distance modulus $\mu$, while the red curve shows
Gaussian with the same mean and standard deviation as the computed pdf. The
green curve shows the pdf which is a Gaussian in distance and has the mean and
standard deviation of the computed pdf in distance, while the blue curve shows the
pdf which is a Gaussian in parallax and has the mean and dispersion of the
computed pdf in parallax. None of the coloured curves can be considered a reasonable
representation of the computed pdf. The clear message of \figref{fig:badpdfs}
is that it is dangerous to quantify the distance to these stars in the form
$x\pm y\kpc$ because this notation implies that a Gaussian pdf adequately
approximates the true pdf.

We have derived multi-Gaussian approximations to the pdf in $\mu$ since this
variable is physically meaningful for any real number. We write
 \[\label{eq:defsfk}
P(\mu) = \sum_{k=1}^N {f_k\over \sqrt{2\pi\sigma_k^2}}
\exp\bigg(-{(\mu-\mu_k)^2\over2\sigma_k^2}\bigg), 
\]
 where $N$, the means $\mu_k$, weights $f_k$, and dispersions $\sigma_i$ are to be
determined. We take bins in distance modulus of width $w_i = 0.2$,
containing a fraction $p_i$ of the total probability taken from the
computed pdf, and a fraction $P_i$ of the total probability taken from the
multi-Gaussian approximation and consider the statistic
\[\label{eq:defsF}
F = \sum_i \left(\frac{p_i}{w_i}-\frac{P_i}{w_i}\right)^2\tilde{\sigma} w_i
\]
where the weighted dispersion
\[
\tilde{\sigma}^2 \equiv \sum_{k=1,N} f_k \sigma_k^2
\]
 is a measure of the overall width of the pdf.  Our definition of $F$
includes the factor $\tilde{\sigma}$ to ensure that $F$ is unchanged when the
width of both the true pdf and our approximation are increased by the same
factor: this condition ensures that $F$ is a measure of how well the shape of
the distribution is fitted. We use $\tilde{\sigma}$ in equation
(\ref{eq:defsF}) rather than the dispersion of the pdf because in some
circumstances (double or triple peaked distributions) the dispersion is
dominated by the distance between peaks, rather than the widths of the
individual peaks themselves, and it is the peaks that should set the scale. A
practical difficulty is that $F$ is minimised by letting every
$\sigma_k\rightarrow0$. Hence instead of minimising $F$, we minimise the
alternative statistic
 \[
F' = \sum_i \left(\frac{p_i}{w_i}-\frac{P_i}{w_i}\right)^2\; w_i
\]
 and only use $F$ to measure whether the fit is a sufficiently accurate
description of the data.

\begin{figure*}
\centerline{\epsfig{file=pjm/hotH.ps,width=.3\hsize}
\epsfig{file=pjm/coolH.ps,width=.3\hsize}
\epsfig{file=pjm/giantH.ps,width=.3\hsize}}
\caption{Histograms of the difference between spectrophotometric
and Hipparcos parallaxes. The left panel is for hot dwarfs, the centre panel
is for cool dwarfs and the right panel is for giants. The full curves
are Gaussians of zero mean and unit dispersion, not fits to the data. The
black points are obtained from the simple expectation of $\varpi$ while the
red points are obtained as described in the text from the Gaussian
fits to the pdf in distance modulus.
In these and subsequent histograms the vertical axis plots $\d N/\d x$, the
horizontal error bars mark the widths of the bins and the vertical error bars
indicate Poisson uncertainties.}\label{fig:Hipp_pi_e}
\end{figure*}

\begin{figure}
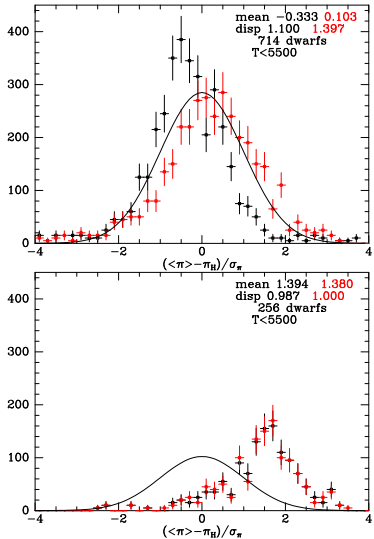

\centerline{\epsfig{file=pjm/cd_many.ps,width=.6\hsize}}
\centerline{\epsfig{file=pjm/cd_one.ps,width=.6\hsize}}
\caption{The centre histogram of \figref{fig:Hipp_pi_e} broken down into cold
dwarfs with single-Gaussian (lower) and multiple-Gaussian (upper)
pdfs.}\label{fig:manyGcds}
\end{figure}

\begin{figure*}
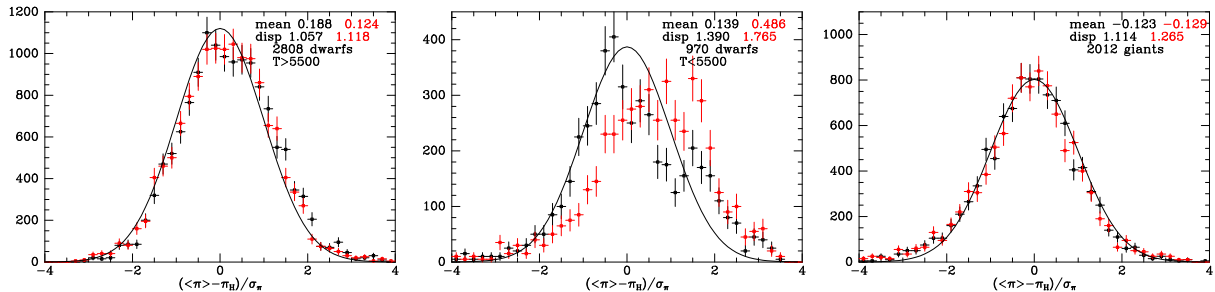

\centerline{\epsfig{file=pjm/hot_noAv.ps,width=.3\hsize}
\epsfig{file=pjm/cool_noAv.ps,width=.3\hsize}
\epsfig{file=pjm/giant_noAv.ps,width=.3\hsize}}
 \caption{Histograms of the difference between the Hipparcos parallaxes and
expectation of the parallax from the spectrophotometry  for hot dwarfs
(left), cool dwarfs (centre) and giants (right) when the extinction
is assumed to be zero. The full curves
are Gaussians of zero mean and unit dispersion.}\label{fig:Hipp_pi2_noEH}
\end{figure*}

\begin{figure*}
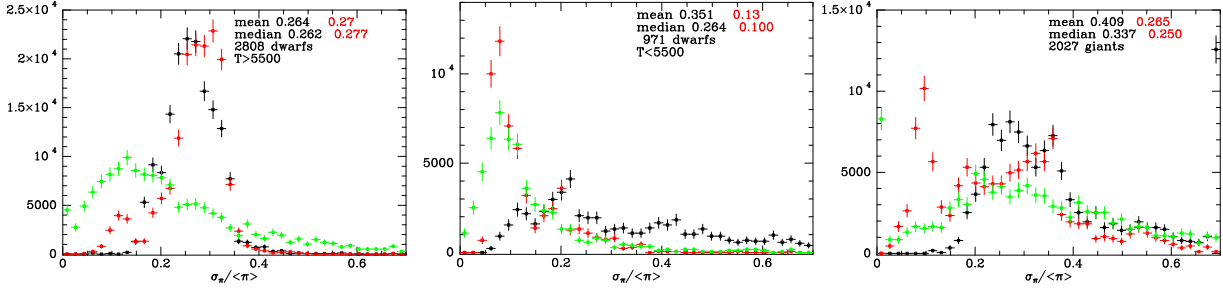

\centerline{\epsfig{file=pjm/hot_piH.ps,width=.3\hsize}
\epsfig{file=pjm/cool_piH.ps,width=.3\hsize}
\epsfig{file=pjm/giant_piH.ps,width=.3\hsize}}
 \caption{The distributions of errors in Hipparcos parallaxes (green) and in our
spectrophotometric parallaxes (black and red) for hot dwarfs
(left), cool dwarfs (centre) and giants (right). The black points show the
errors in $\ex{\varpi}$ computed directly. The red points show the
errors in parallaxes derived from the Gaussian fits to the pdf in distance
modulus.}\label{fig:pipi}
\end{figure*}

If the value of $F$ for a Gaussian with the same mean and dispersion in $\mu$
as that taken from the computed pdf is less than a threshold value $F_t=0.04$, we
accept this as an adequate description of the data. This condition holds for
around 45 per cent of the RAVE stars. When it fails, we use the
Levenberg-Marquardt algorithm to minimise $F'$ with $N=2$ and several
different initial choices for the parameters. We accept this description of
the data if it gives $F<F_t$ \emph{and} the dispersion of the model is within
20 per cent of that of the complete pdf. The latter condition ensures that we do not
accept models that provide an excellent fit to a significant component of
the probability but ignore a small but non-negligible component at a
different distance.  If the two-Gaussian description fails, we fit a
three-Gaussian approximation.  We reach this stage for around 5 per cent of
the RAVE stars because the double-Gaussian approximation is accepted in
$\sim50$ per cent of cases.  \figref{fig:goodpdfs} shows the multi-Gaussian models
fitted to the pdfs shown in \figref{fig:badpdfs}.

Any fits for which the dispersion of the fitted model differs by more than 20
per cent from that of the data is flagged as possibly
inadequate. Approximately four per cent of the models are flagged for
this reason. In \figref{fig:badflag} we show some typical
examples of the flagged models. We see that the problems are in fact minor ones.

\section{Hipparcos stars}\label{sec:Hipp}

As in B11, the primary test of the validity of our spectrophotometric
distances is provided by Hipparcos stars that are likely to be single stars
because in the \cite{vanLeeuwen} catalogue they have ${\tt soln}<10$. There
are 5614 distinct stars of this type for which we have RAVE parameters, and
the mean S/N ratio of their spectra is 84.

The quoted errors on the stellar parameters play a big role in the Bayesian
algorithm, and good results are obtainable only with accurate error
estimates. When the data were first processed using only the internal error
estimates produced by the spectral-reduction pipeline, manifestly
inconsistent results for Hipparcos stars were produced. The results were
dramatically improved by adding to the internal errors the external errors
for various classes of star derived by \cite{DR4} and listed in
Table~\ref{tab:ext}. The quadrature sums of the internal and external errors
prove to be quite similar to the errors adopted by B11, which could not be
founded on star-specific error estimates from the old pipeline.

The black points in \figref{fig:Hipp_pi_e} show histograms of the
discrepancies between Hipparcos parallaxes $\varpi_{\rm H}$ and expectation values of
parallaxes obtained from $P({\rm model|data})$ for three groups of stars:
giants ($\log g<3.5$), hot dwarfs ($\Teff>5500\K$) and cool dwarfs.  The
parallax differences are normalised by the quadrature sum of the formal
errors in the Hipparcos data and our adopted errors, so if our procedure were
sound and the central limit theorem applied to the data, the histograms would
be Gaussians of unit dispersion. This expectation is met to a pleasing extent
for hot dwarfs and giants -- for the hot dwarfs the mean of the distribution
is $0.143$ and the dispersion is $1.061$ and for the giants they are $-0.057$
and $1.077$.  Thus on average the parallaxes of the hot dwarfs are slightly
too large, while those of the giants are slightly too small and our error
estimates are only a shade too small. The results for the smaller number of
cool dwarfs are less clear-cut: the mean and dispersion are $0.123$ and
$1.314$ implying that our parallaxes are slightly too large and our errors
are materially too small.

\begin{table}
\caption{Estimates of the external errors in the stellar parameters.
the boundary between ``metal-poor'' and ``metal-rich'' lies at
$\hbox{[M/H]}=-0.5$, and between ``hot'' and ``cool'' lies at $6000\K$.}
\label{tab:ext}
\begin{center}
\begin{tabular}{lcccc}
\hline
stellar type&N&$\sigma(\Teff)$&$\sigma(\log g)$&$\sigma(\hbox{[M/H])}$\\
\hline
\multispan{5}{\hfil dwarfs\hfil}\\
\hline
hot, metal-poor&          28&      314&     0.466&     0.269\\
hot, metal-rich &        104&      173&     0.276&     0.119\\
cool, metal-poor&         97&      253&     0.470&     0.197\\
cool, metal-rich&        138&      145&     0.384&     0.111\\
\hline
\multispan{5}{\hfil Giants\hfil}\\
\hline
hot      &          8&      263&     0.423&     0.300\\
cool, metal-poor&        273&      191&     0.725&     0.217\\
cool, metal-rich&        136&       89&     0.605&     0.144\\
\hline
\end{tabular}
\end{center}
\end{table}

\begin{table}
\caption{Mean distance ratios for Hipparcos stars. Ideally all entries would
be unity.}\label{tab:rSD}
\begin{center}
\begin{tabular}{lcccc}
&$\overline{\ex{s}/s_{\ex{\mu}}}$&$\overline{s_{\ex{\mu}}\ex{\varpi}}$&
$\overline{\ex{\varpi}/\varpi_{\rm H}}$&$\overline{\ex{s}\varpi_{\rm H}}$\\
\hline
Hot dwarfs	&1.045&1.040&0.958&1.042\\
Cool dwarfs	&1.116&1.094&1.132&1.447\\
Giants		&1.111&1.093&1.115&1.386\\
\hline
\end{tabular}
\end{center}
\end{table}

It is interesting to compute means  of the distances  ratios. Let
 \begin{eqnarray}
r_{s\mu}&\equiv&\overline{\ex{s}/s_{\ex{\mu}}}\quad
r_{\mu\varpi}\equiv \overline{s_{\ex{\mu}}\ex{\varpi}}\nonumber\\
r_{\varpi{\rm H}}&\equiv&\overline{\varpi_{\rm H}/\ex{\varpi}}\quad
r_{s{\rm H}}\equiv\overline{\ex{s}\varpi_{\rm H}},
\end{eqnarray}
 where overbars imply averages of a group of stars and $s_{\ex{\mu}}$ is the
distance implied by the expectation value of the distance modulus.
Table~\ref{tab:rSD} gives these ratios for hot dwarfs, cool dwarfs and giants.
For the hot dwarfs all ratios are pleasingly close to unity, but for both the
cool dwarfs and the giants we see that $\ex{s}$ gives a systematically larger
distance than $s_{\ex{\mu}}$, which in turn gives a bigger distance than
$1/\ex{\varpi}$, which itself gives a bigger distance than $1/\varpi_{\rm
H}$, which we take to be the most reliable distance estimator. These biases
are easily understood in terms of the weights that each estimator attaches to
possibilities of long or short distances. The comparisons with the Hipparcos
parallaxes clearly indicates that for stars with wide distance pdfs (cool
dwarfs and giants), $1/\ex{\varpi}$ performs much better than either
$\ex{s}$ or $s_{\ex{\mu}}$.

The red points \figref{fig:Hipp_pi_e} show histograms of discrepancies
between the Hipparcos parallaxes and parallaxes based on the multi-Gaussian
fits to the distance moduli as follows. When a single Gaussian has been
fitted, we convert the mean and dispersion of this Gaussian into a parallax
and its error by standard formulae. If two or three Gaussians have been
fitted, we choose the Gaussian that makes the Hipparcos parallax most
probable and convert the mean and dispersion of this Gaussian to a parallax
and its error as before. The red histogram for the hot dwarfs is an almost
perfect realisation of the unit Gaussian while that for the giants is only
marginally less satisfactory than the corresponding black histogram.  The red
histogram for the cool dwarfs is both significantly displaced to the right and
broader than it should be.  

\figref{fig:manyGcds} clarifies the situation by splitting the histogram of
the cool dwarfs into those with pdfs that have been fitted with a single
Gaussian (lower panel) and those with multi-Gaussian fits (upper panel). We
see that for the latter stars the crude mean of possible parallaxes is
smaller than it should be, and a more satisfactory distribution of
spectrophotometric parallaxes is obtained if Hipparcos is used to choose
between the Gaussians.  The lower panel in \figref{fig:manyGcds} shows that
when a cool dwarf has a single-Gaussian pdf, its parallax is systematically
over-estimated. When the single- and multi-Gaussian samples are aggregated in
\figref{fig:Hipp_pi_e}, the over-estimated parallaxes of the single-Gaussian
stars combine with the under-estimated parallaxes of the multi-Gaussian stars
to produce a deceptively satisfactory black histogram. The mean S/N ratio of
the Hipparcos stars with single-Gaussian fits is lower than that of the stars with
multi-Gaussian fits (51.0 versus 66.5), so one suspects that with poorer data
the system loses track of the possibility that the star has left the main
sequence.

We test the soundness of the probabilities assigned to each Gaussian
component of the pdf by calculating the sums $s_k=\sum_{\rm stars}1/f_k$,
where $k=1,2,3$ depending on which Gaussian component the Hipparcos data
points to, and $f_k$ is the weight of that component. Given a large and
sample of stars with accurate parallaxes (so the true component is always
chosen), $s_k$ should be independent of $k$ because when $f_k$ is small, that
component will be rarely chosen so $s_k$ will have a small number of large
contributions, while a component with large $f_k$ will be chosen often, but
each contribution to $s_k$ will be modest. When we compute mean values of
$1/f_k$ for our Hipparcos stars, we find 441/2807 hot dwarfs with two
Gaussians fitted, and for these stars we find $s_k=(444,458)$. Similarly,
615/970 cool dwarfs have two Gaussians and for these stars we find
$s_k=(577,2100)$, while 100 cool dwarfs have three Gaussians and for these
stars $s_k=(94,126,476)$.  934/2015 giants have two Gaussians and these stars
yield $s_k=(779,3593)$ while 492 giants have three Gaussians and for these
stars $s_k=(350,759,748)$. These results suggest that the probabilities
assigned the various Gaussians are broadly correct although there is a
tendency for too little probability to be assigned to the weakest components.

The likely explanation of the neglect of weaker components is that the
Hipparcos stars are biased towards nearer stars because stars thought to be
near, usually on account of having large proper motions, preferentially
entered the Hipparcos Input Catalogue. Consequently, we have tested the
constancy of the $s_k$ for a sample in which distant options will have been
rather rarely chosen. For the giants the distant option is the more probable
one, so it is natural that for these stars Hipparcos chooses the less
probable Gaussian more often than one would expect if we had parallaxes for
every star in our sample.

\figref{fig:Hipp_pi2_noEH} shows the effect of setting $A_V=0$ for all stars.
With reddening neglected, dwarfs must be moved to lower masses to match the
observed colours, and the consequent diminution of their luminosities causes
them to be brought closer to match the observed magnitudes. The overall
effect is to increase the spectrophotometric parallaxes of hot dwarfs by
$\sim0.05\sigma$, so those of the hot dwarfs are now on average too large by
$\sim0.19\sigma$, while those of the cool dwarfs are too large by
$\sim0.14\sigma$. With extinction neglected, giants need to be moved away to
diminish their brightnesses so their histogram of $\ex{\varpi}-\varpi_{\rm
H}$ moves leftward, and our parallaxes become too small by $0.12\sigma$ on
average. Thus the Hipparcos stars convincingly validate our procedure for
taking into account the effects of dust.

\figref{fig:pipi} compares the distribution in the fractional errors in
Hipparcos parallaxes (shown in green) with the corresponding errors in our
parallaxes: the black points are for the straightforward expectation values
of $\varpi$ while the red points are for the parallaxes computed from the
multi-Gaussian fits to the pdfs in distance modulus.  For hot dwarfs the
black and red histograms are similar because few of these stars
have multi-modal pdfs. They show error distributions that are materially
narrower than that from Hipparcos, with most values of
$\sigma_\pi/\ex{\varpi}$ falling in the range $(0.18,0.38)$ with a median
value of $0.26$. 

For the cool dwarfs the black and red histograms are quite different in that
the red histogram shows a substantial population with spectrophotometric
parallaxes in error by less that 10\% and essentially no stars with errors
greater than 35\%.  The stars with $\sigma_\varpi/\ex{\varpi}<0.1$ are stars
that the spectrophotometry cannot securely assign to dwarfs or giants until
astrometric data become available -- in the present case a Hipparcos
parallax. There will probably be many stars of this type in the Gaia
Catalogue. The red histogram for the giants shows a similar if smaller
population of stars.

For now we must live with dwarf/giant confusion and the black histograms of
parallax errors are most relevant. These show that the spectrophotometric
parallaxes of cool dwarfs are not competitive with Hipparcos parallaxes, in
contrast to the case of some hot dwarfs and a number of giants, which do have
more precise spectrophotometric parallaxes than Hipparcos parallaxes.  Thus
the competitiveness of the spectrophotometric parallaxes vis a vis Hipparcos
parallaxes increases along the sequence cool dwarfs to hot dwarfs to giants
in parallel with the increase in the luminosities and thus typical distances
of these stars.

\begin{table}
\caption{Ratios of distance measures for general stars with $s<2\kpc$}\label{tab:gen_ratios}
\begin{tabular}{lcccc}
&$N_\star$&$\overline{\ex{s}/s_{\ex{\mu}}}$&$\overline{s_{\ex{\mu}}\ex{\varpi}}$&$\overline{\ex{s}\ex{\varpi}}$\\
\hline
&\multispan3{Giants ($\log g<3.5$)\hfil}\\
\hline
$\log g>2.4$		&69008	&1.11&1.13&1.26\\
Red Clump		&39900	&1.04&1.04&1.09\\
$\log g<1.7$		&28472	&1.06&1.05&1.11\\
\hline
&\multispan3{Dwarfs ($\log g\ge3.5$)\hfil}\\
\hline
$\Teff>6500$		&22701	&1.04&1.03&1.07\\
$5500<\Teff\le6500$	&71641	&1.04&1.04&1.08\\
$5200<\Teff\le5500$	&19697	&1.08&1.08&1.17\\
$\Teff\le5200$		&27408	&1.13&1.12&1.29\\
\hline
\end{tabular}
\end{table}

\section{Distances to all stars}\label{sec:all}

We have examined the statistics of distances to RAVE stars as functions of a
cutoff in the S/N of the analysed spectrum and found that dependence on the
cutoff S/N is weak.  Below we report results obtained for stars with
$\hbox{S/N}\ge10$ -- the mean S/N ratio for such stars that lie closer than
$1.3\kpc$ is 33.

We have investigated the sensitivity of our distances to the model of the
disc used in the prior (eqs \ref{eq:thindisc} and \ref{eq:thickdisc}) by
re-evaluating the distances to every twentieth star in the catalogue with the
scale radii and scale heights of both discs multiplied by a factor $1.5$. The
resulting histogram of ratios $\ex{\varpi}_2/\ex{\varpi}_1$ of the parallax
with the revised prior to the parallax with the standard prior peaks sharply
at $1.02$ but has a long tail to values $\sim1.2$ with the consequence that
the mean of this ratio is $1.045$. This result shows that, as one would hope,
our results are not sensitive to the prior.

Table~\ref{tab:gen_ratios} shows the ratios of the available distance
measures for ordinary stars, broken down into giants and dwarfs,
with the giants subdivided into stars with lower surface gravity than the red
clump ($1.7<\log g<2.4$ and $0.55\le J-K\le0.8$), the red clump itself and
stars with higher gravities.  We see that in every case the distances are
ordered $\ex{s}>s_{\ex{\mu}}>1/\ex{\varpi}$. Moreover, $\ex{s}$ and $1/\ex{\varpi}$
are discrepant at the 26\% level for the highest-gravity giants and coolest
dwarfs, while for moderately cool dwarfs these measures are discrepant at the
17\% level. 

\subsection{Kinematic distance corrections}\label{sec:SBA}

Sch\"onrich et al. (2012; hereafter SBA) describe a technique that uses the
kinematics of stellar populations to identify and correct systematic errors
in distances, and we can use this technique to determine which of our
discrepant distance estimates is most reliable, and potentially to correct
the most reliable measure for any systematic bias.

The corrections of SBA are based on the assumption that one knows roughly how
the velocity ellipsoid is oriented at each point in the Galaxy, and that the
only mean-streaming motion is azimuthal circulation at a speed $v(R,z)=\Theta
g(R,z)$, where $\Theta$ is an unspecified constant and $g(R,z)$ is a function
one chooses. We adopt
 \[
g=\sqrt{1-(2\psi/\pi)^2}, \hbox{ where }\psi\equiv\arctan(z/R),
\]
 which has an appropriate form, but the results are very insensitive to the
choice of $g$: essentially unchanged results are obtained with $g=1$.  The
algorithm involves converting heliocentric velocities to Galactocentric
velocities and thus requires assumptions regarding the Galactocentric
velocity of the Sun and the distance $R_0$ to the Galactic centre. We assume
that $R_0=8.33\kpc$, that the local circular speed is $\Theta_0=230\kms$, and
that the Sun's velocity with respect to the Local Standard of Rest is
$(U_0,V_0,W_0)=(11.1,12.24,7.25)\kms$ \citep{SchoenrichBD}. There is very
little sensitivity to the value of $\Theta_0$. The azimuthal direction is
assumed to be a principal axis of the velocity ellipsoid, while the latter's
longest axis is tilted with respect to the plane by angle
$\beta=a_0\arctan(z/R)$, where $a_0$ is a parameter. 

The corrections exploit pattern on the sky of correlations between the local
Cartesian velocity components $U,V,W$ that are introduced by distance errors.
To assess the magnitude of these correlations one has first to correct the
raw correlations for contributions from sources other than distance errors.
The most important such source is observational errors in the proper motions,
so knowledge of the magnitude of these errors is needed for the correction.

\begin{figure}
\centerline{\epsfig{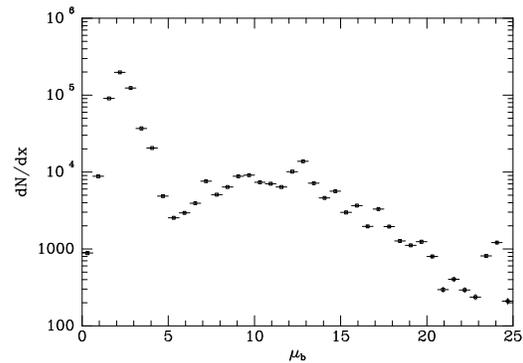}}
\caption{Histogram of the errors in the component $\mu_b$ of proper motion for
all RAVE stars. The long tail of errors in excess of $8\mas\yr^{-1}$ gives
rise to nonsensical results, so we drop stars with such large proper-motion
errors from the sample.}\label{fig:pmerrs}
\end{figure}

Proper motions for RAVE stars can be drawn from several catalogues.
\cite{Williams13} compares results obtained with different proper-motion
catalogues, and on the basis of this discussion we decided to work with the
PPMX proper motions \citep{Roeseretal} because these are available for all our
stars and they tend to minimise anomalous streaming motions.
\figref{fig:pmerrs} shows a histogram of the errors for RAVE stars given in
the PPMX catalogue. It shows that there is a fat tail in the error
distribution, and one may show that this tail should not to be taken at face
value because when one calculates the velocity dispersions of all the RAVE
stars in spatial bins that are further than $\sim0.5\kpc$ from the Sun, the
dispersions are often smaller than the contribution expected from
proper-motion errors alone. This paradox disappears if one cuts stars with
errors in one component of proper motion greater than $8\mas\yr^{-1}$, and we
impose this cut throughout the SBA analysis. The only class of stars that is
significantly depleted by this cut is that of the very cool dwarfs, which
shrinks from $38\,330$ stars to $27\,332$ stars.  This cut reduces the rms
error in one component of proper motion to $2.5\mas\yr^{-1}$.  

A second source of correlations that complicate the SBA analysis is rotation
of the velocity ellipsoid's principal axes as one moves around the Galaxy,
and a model of the velocity ellipsoid is used to correct for this effect.
The final product is the factor $1+f$ by which all distances must be
contracted (or expanded if $f<0$) for all correlations between $U$, $V$ and
$W$ to be accounted for by a combination of observational errors and rotation
of the principal axes of the velocity ellipsoid.

SBA give two formulae for corrections, one, $f_U$, involving ``targeting''
$U$ and one, $f_W$, using $W$ as a target. Because the latter is independent of
azimuthal streaming, it is the simpler and more reliable. Their equations
(19) and (38) give the $W$ and $U$ correction factors, respectively,
after the raw covariances have been corrected for observational errors using
their equations (22) and (25). 

\begin{table}
\caption{Kinematic correction factors for general stars at $s<2\kpc$. The
first two columns give results of a test in which all stars were recorded to
be further from the Sun than their true locations by a factor $1.3$. The
last two columns are computed from the real RAVE catalogue.}\label{tab:tests}
\begin{tabular}{lcccc}
&$f_W(T)$&$f_U(T)$&$f_W$&$f_U$\\
\hline
&\multispan3{Giants ($\log g<3.5$)\hfil}\\
\hline
$\log g>2.4$		&0.304	&0.323	&0.134	&0.248\\
Red Clump		&0.311	&0.332	&0.160	&0.249\\
$\log g<1.7$		&0.310	&0.348	&0.453	&0.676\\
\hline
&\multispan3{Dwarfs ($\log g\ge3.5$)\hfil}\\
\hline
$\Teff>6500$		&0.295	&0.295	&-0.270	&-0.210\\
$5500<\Teff\le6500$	&0.312	&0.312	&-0.081	&-0.037\\
$5200<\Teff\le5500$	&0.286	&0.286	&-0.064	&-0.027\\
$\Teff\le5200$		&0.306	&0.306	&-0.026	&0.043\\
\hline
\end{tabular}
\end{table}

\begin{table}
\caption{Kinematic correction factors for general stars at $s<1.3\kpc$. The
first two columns report results from a test in which the recorded locations
of stars were further than their true locations by a factor $1+f$ where $f$
is a random variable with mean and dispersion $0.2$.}\label{tab:tests2}
\begin{tabular}{lcccc}
&$f_W(T)$&$f_U(T)$&$f_W$&$f_U$\\
\hline
&\multispan3{Giants ($\log g<3.5$)\hfil}\\
\hline
$\log g>2.4$		&0.203	&0.203	&0.066	&0.185\\
Red Clump		&0.157	&0.157	&0.114	&0.148\\
$\log g<1.7$		&0.100	&0.130	&0.210	&0.334\\
\hline
&\multispan3{Dwarfs ($\log g\ge3.5$)\hfil}\\
\hline
$\Teff>6500$		&0.220	&0.207	&-0.270	&-0.210\\
$5500<\Teff\le6500$	&0.217	&0.217	&-0.081	&-0.037\\
$5200<\Teff\le5500$	&0.227	&0.247	&-0.064	&-0.027\\
$\Teff\le5200$		&0.217	&0.217	&-0.050	&0.041\\
\hline
\end{tabular}
\end{table}

From the RAVE data we have extracted correction factors to the distance
estimator $1/\ex{\varpi}$ for the three types of giants and four types of
dwarf listed in Table~\ref{tab:gen_ratios}.  The code used to determine the
corrections was tested as follows. For each star in a class, the measured
$U,V,W$ velocities were replaced by values chosen from a triaxial Gaussian
velocity ellipsoid that has dispersions $\sigma_i=(40,40/\surd2,30)\kms$
around systematic rotation at $200\kms$.  Most tests were run with the
orientation of the principal axes determined by setting $a_0=0.8$, but
excellent results are obtained with other plausible values of $a_0$,
including zero.  Likewise, the outcome of the code tests is not sensitive to
the adopted dispersions $\sigma_i$. Next proper motions and line-of-sight
velocities are calculated from the model velocities, and Gaussian
observational errors are added with the dispersions that are given in the
PPMX catalogue. Then the stars are moved along their lines of sight
to points more distant by a factor $1+f$ and their $U,V,W$ components are
re-evaluated from the proper motions.  In this way we obtain a catalogue of
phase-space positions for a population of objects whose distances have been
over-estimated by a factor $1+f$. The SBA algorithms are then used to infer
from this catalogue the value of $f$.

The first two numerical columns of Table~\ref{tab:tests} show the fractional distance
excesses $f_W$ and $f_U$ obtained by targeting $W$ and $U$ when distances to
the stars have been over-estimated by a factor $1+f$ with $f=0.3$.
Consequently, ideally we would have $f_W=f_U=0.3$ for all star classes. For
$f_W$ this expectation is borne out for all classes to better than $5\%$, and
for the dwarfs it is similarly for $f_U$. For the giants $f_U$ is up to
$16\%$ larger than it should be, a result which reflects the breakdown of the
approximations made by SBA when dealing with more distant stars.

The final two columns of Table~\ref{tab:tests} show the fractional distance
excesses $f_W$ and $f_U$ for the seven classes of RAVE stars using the
measured distances and velocities when $1/\ex{\varpi}$ is used as the
distance measure. For the giants the differences between $f_W$ and $f_U$ are
in the same sense ($f_U>f_W$) as in the tests but they are larger than in the
tests.  The cause of this difference is not obvious, but one suspects a major
contributor is the well-known existence of clumps of stars in the $(U,V)$
plane \citep{Dehnen98,Famaey,Antoja}, which conflict with the assumption of
simple azimuthal streaming that is fundamental to SBA's derivation of the
formula for $f_U$. Since prominent clumps are absent from the distribution of
Hipparcos stars in the $(U,W)$ and $(V,W)$ planes \citep{Dehnen98}, $f_W$ is
expected to be a more reliable diagnostic of distance errors than $f_U$.
Table~\ref{tab:tests} then suggests that $1/\ex{\varpi}$ over-estimates
distances to high-gravity giants and red-clump stars by $\sim15\%$, and gives
distances to dwarfs that are too small by factors that rise from $\sim5\%$ at
the cool end rising to $\sim25\%$ at the hot end.

In selecting stars for inclusion in the SBA analysis we have imposed a limit
$s_{\rm max}$ on the reported distance, and the results one obtains for both
the test and with the real data depend on the value chosen for $s_{\rm max}$.
Table~\ref{tab:tests} is based on the choice $s_{\rm max}=2\kpc$.
Table~\ref{tab:tests2} is based on $s_{\rm max}=1.3\kpc$ and the results of
tests reported in the first two numerical columns of this table differ from
those reported in the corresponding columns of Table~\ref{tab:tests} in that
the distances to stars were increased by a factor $1+f$ where $f$ is now a
Gaussian random variable with mean and dispersion $0.2$. The test results are
fairly satisfactory for the dwarfs in that both $f_W$ and $f_U$ have values
within $\sim10\%$ of the true value, $0.2$. The test results for the giants
are decidedly less satisfactory in that the $f$ values are too small by an
amount which increases with the typical luminosity within a class. It is easy
to understand why this is so: stars that happen to get a large fractional
distance increase are liable to be pushed beyond $s_{\rm max}$ whilst stars
that have their distances decreased can enter the sample from beyond $s_{\rm
max}$, and the SBA algorithm correctly infers that on average the stars
\emph{in the analysed sample} have small distance over-estimates even though
in the population as a whole stars have larger distance over-estimates.
Clearly, for this phenomenon to be important the catalogue needs to contain
many stars that really are at distances $\sim s_{\rm max}$.  The dwarfs do
not satisfy this condition, but the low-gravity giants very much straddle the
$1.3\kpc$ distance cut. 

Comparing columns 3 and 4 of Table~\ref{tab:tests2} with the corresponding
column of Table~\ref{tab:tests} we see that reducing $s_{\rm max}$ from
$2\kpc$ to $1.3\kpc$ has only a modest effect on the $f$ values for dwarfs
and a significant effect on giants.  The $f$ values of giants decrease
significantly for all three classes, but the final $f_W$ factors still
increase with decreasing gravity contrary to the tendency seen in the test,
so we really must be over-estimating distances to the lowest-gravity (and
most luminous) giants. A possible explanation is that we are using stellar
parameters obtained under the assumption of Local Thermodynamic Equilibrium
(LTE). The validity of LTE decreases with $\log g$, and when non-LTE effects
are taken into account, the recovered gravity of a giant star increases
\citep{Ruchti}, and the predicted luminosity decreases, bringing the star
closer. 

\begin{figure*}
\centerline{\epsfig{file=pjm/giant_sfac.ps,width=.45\hsize}\ 
\epsfig{file=pjm/dwarf_sfac.ps,width=.45\hsize}}
\caption{The squares show the ratios $1+f_W$ of the
spectrophotometric distances $1/\ex{\varpi}$ to the true distances for giants
with $1/\ex{\varpi}<1.3\kpc$
broken down by $\log g$ (left) and dwarfs broken down by $\Teff$ (right).
All bins contain the same number of stars and each star is in only one bin so
the data points are mutually independent.  The
triangles show the same quantities inferred for Hipparcos stars from their
directly measured parallaxes, again using equally populated and independent
bins. The red curve is fifth-order Chebyshev polynomials fitted
to all the data.}\label{fig:dist_fac}
\end{figure*}

\begin{figure}
\centerline{\epsfig{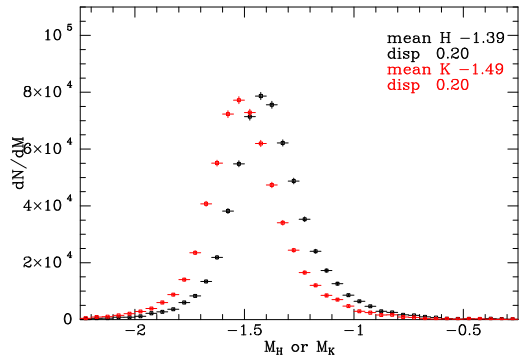}}
\caption{The distribution of absolute $H$ and $K$ magnitudes of red-clump
stars with $1/\ex{\varpi}<1.3\kpc$.}\label{fig:rclump}
\end{figure}

The squares in \figref{fig:dist_fac} show the values of $1+f_W$ obtained when
the giants are grouped by $\log g$ and the dwarfs are grouped by $\Teff$ --
in each case the SBA algorithm is used on 15 bins of stars at
$1/\ex{\varpi}<1.3\kpc$ with equal numbers of stars in each bin, and all bins
statistically independent. The triangles show the analogous ratios
$\varpi_{\rm H}/\ex{\varpi}$ of our distance to that implied by the Hipparcos
parallax. The curves show fifth-order polynomial fits to all the points. The
squares and triangles tell the same story from a qualitative perspective:
along the sequence of giants there is a steady increase in the tendency to
over-estimate distances as one moves to lower gravity (and higher
luminosity), while the dwarfs show a clear trend towards distance
over-estimation with falling $\Teff$ with the exception of the coolest bin,
which shows marked distance under-estimates. The SBA points for dwarfs tend
to lie below those from Hipparcos, so SBA and Hipparcos disagree about the
value of $\Teff$ at which our distances are unbiased.

Our tests suggest that $f_W$ should be a reliable guide to any systematic
errors in the distances to our dwarf stars. The situation regarding the giant
stars is less clear because the $f$ values are biased low unless $s_{\rm
max}$ is large enough to encompass most of the stars in the catalogue.
Unfortunately, the more distant stars are, the more sensitive the returned
value of $f_U$ becomes to restrictive assumptions regarding the pattern of
mean-streaming and random velocities in the Galaxy and some approximations.
The value of $f_W$ is less sensitive to these issues and therefore more
reliable, but its sensitivity to $s_{\rm max}$ is worrying.  A further blow
to the credibility of $f_W$ will emerge below from an analysis of the
red-clump stars.

\subsubsection{Kinematic corrections to multi-Gaussian pdfs}

SBA assume that one is working with a simple distance estimator, while in
Section \ref{sec:pdfs} we saw that our most complete information is contained
in a distance pdf. Can we use a kinematic analysis to refine these pdfs?

The SBA algorithm involves several sample averages such as $\ex{Wy}$, where
$W$ and $y$ are quantities that depend on the distance to each star. In our
analysis above we evaluated these for just one  distance, but given a pdf
$P(\mu)$ it is
straightforward  to replace $Wy$ by the expectation value of $Wy$:
 \[
\overline{Wy}\equiv\int\d\mu\,P(\mu)W(\mu)y(\mu).
\]
 These expectation values are then averaged over the sample to produce the
sample averages $\ex{Wy}$, etc., that appear in the SBA formalism. Thus is
straightforward to use the pdfs to calculate a kinematic correction factor
such as $f_y$.

It is less clear how one should modify the pdf in light of a non-zero value
of $f_y$. We have experimented with two possibilities.

\begin{itemize}
\item[(i)]  Move the centres of all the Gaussians to larger
or smaller distance moduli until, $f_y=0$. This procedure produces results
that are rather similar to, but slightly less convincing than, those obtained
without the pdfs.

\item[(ii)] When a star has more than one Gaussian in its pdf, modify the
probabilities $f_k$ (eqn.~\ref{eq:defsfk})
associated with the two most probable Gaussians. This procedure
is appropriate if the Bayesian algorithm has correctly identified the  two
model stars that an observed star could be, but, perhaps driven by a faulty
prior, has assigned inappropriate odds to the options. We now report results
obtained with this procedure.

\end{itemize}

We make the probabilities $f_1$ and $f_2$ in equation (\ref{eq:defsfk}) a
function of a variable $\theta$ through
\[\label{eq:defstheta}
f_1=A\cos^2(\theta),\quad f_2=A\sin^2(\theta),
\]
 where at the outset we fix $A\equiv f_1+f_2$ to be the total probability
associated with the two most probable options. Then we make $\theta$, which
is confined to the range $(0,\pi/2)$, a function of a variable $\xi$ that
can span the whole real line, through
 \[\label{eq:defsxi}
\theta=\arctan(\e^\xi).
\]
 The original values of $f_i$ determine starting values for $\theta$ and
$\xi$. If the kinematic analysis has returned $f_y>0$, implying that
distances need to be shortened and the first Gaussian describes a nearer
option than the second, then we lower $\theta$ by subtracting $5f_y$ from
$\xi$ -- the factor 5 is arbitrary: smaller values lead to slower convergence
of the iterations but larger values can cause the iterations to undergo
diverging oscillations. If, conversely, $f_y<0$, we need to increase $\theta$
and $\xi$ so we add $5f_y$ to $\xi$.

\begin{table}
\caption{Kinematic corrections to the pdfs. A large average value of the
parameter $\xi$ defined by equation (\ref{eq:defsxi}) implies that all
the probability has been drive into one Gaussian. The second and third
numerical
columns give the initial and final values of the kinematic error estimator,
which is ideally zero.}\label{tab:pdfsba}
\begin{tabular}{lccc}
&$\overline{|\xi|}$&$f_y(\i)$&$f_y({\rm f})$\\
\hline
&\multispan2{Giants ($\log g<3.5$)\hfil}\\
\hline
$\log g>2.4$		&2.16	&0.403	&0.004\\
Red Clump		&24.85	&0.906	&0.858\\
$\log g<1.7$		&5.55	&0.290	&0.134\\
\hline
&\multispan2{Dwarfs ($\log g\ge3.5$)\hfil}\\
\hline
$\Teff>6500$		&8.22	&-0.252	&-0.247\\
$5500<\Teff\le6500$	&0.39	&-0.015	&-0.009\\
$5200<\Teff\le5500$	&0.55	&0.030	&0.004\\
$\Teff\le5200$		&1.99	&0.247	&0.005\\
\hline
\end{tabular}
\end{table}

In Table~\ref{tab:pdfsba} shows results obtained by iterating up to six times
or until $|f_y|<0.01$. The first numerical column gives the mean of $|\xi|$
for all stars that have more than one Gaussian. A value greater than $\sim3$
implies that all available probability has been driven into whichever
Gaussian will reduce $|f_y|$. For the giants this condition is reached after
about four iterations and is signalled by successive values of $f_y$ becoming
nearly identical.  The second column gives the initial value of $f_y$ and the
third column gives the value of $f_y$ at the end of the iterations. We see
that in the case of the highest-gravity giants, adjusting the $f_i$ has
reduced $f_y$ to the target value, but that there is insufficient ambiguity
in the nature of the clump stars and the low-gravity giants to get $f_W$
below the target value.

There is very little ambiguity in the nature of the hottest dwarfs, so the
procedure makes no significant progress in eliminating the tendency for their
distances to be under-estimated.

The procedure succeeds with the remaining dwarfs: for all three classes
$|f_y|$ is reduced to below the target value, and the modest values of
$\overline{|\xi|}$ given in the first numerical column show that this is
achieved without driving all the probability into one option.

From this analysis we conclude that there is sufficient ambiguity in the
nature of stars that are cooler that $\Teff=5500\K$ and have $\log g>2.4$ to
account for non-zero values of the SBA factor $f_W$ but too little ambiguity
in the nature of hotter dwarfs and low-gravity giants to account for non-zero
$f_W$.

\subsection{Absolute magnitude of the red clump}

Helium-burning stars in the red clump have frequently been used as standard
candles \citep[e.g.][]{Cannon70,Pietrzynski}. Recently \cite{Williams13} used
clump stars in the RAVE survey to analyse the velocity field around the Sun,
and reviewed our knowledge of the absolute magnitudes of these objects and
the possibility that they depend on age and metallicity. They identified
$78\,019$ clump stars as those satisfying the cuts $0.55\le J-K\le0.8$ and
$1.8\le\log g\le3$, where $\log g$ was taken from the vDR3 pipeline
\citep{DR3}. We use the same colour range but a narrower band $(1.7,2.4)$ in
$\log g$ and with gravity taken from the vDR4 pipeline \citep{DR4}.

\figref{fig:rclump} shows the distributions of $H$- and $K$-band absolute
magnitudes for distance $1/\ex{\varpi}$ of clump stars. The distributions are
satisfyingly narrow -- each has a standard deviation of $0.20\mag$ -- but
they are skew, so while their means lie at $M_H=-1.39$ and $M_K=-1.49$ their
peaks lie at $M_H=-1.42$ and $M_K=-1.53$.  These magnitudes are in the SAAO
system: using the formulae of \cite{Koen07} to convert to the 2MASS system we
find the mean of $M_K$ to be $M_K=-1.51$. The sample was restricted by
$1/\ex{\varpi}<1.3\kpc$ but increasing the distance cutoff to $2\kpc$ only
changes the mean absolute magnitudes to $M_H=-1.36$ and $M_K=-1.46$.  For
comparison \cite{Laney12} determined $M_H=-1.49\pm0.022$ and
$M_K=-1.61\pm0.022$ from a sample of 191 Hipparcos stars, and
\citep{Williams13} used calibrations in which the 2MASS absolute magnitudes
were $M_K=-1.65$, $-1.54$ and $-1.64+0.0625z/\hbox{kpc}$. In the last
calibration the decrease in luminosity with increasing distance from the
plane reflects the expected increasing age and decreasing metallicity of
clump stars. However, the age-metallicity sensitivity of the absolute
magnitude is expected to be smallest in the K band \citep[e.g.][]{Salaris}.
Several issues require discussion when considering why our values are
$\sim0.1\,$mag fainter than those of Laney et al.

\begin{itemize}

\item One might argue that the figures given above actually under-estimate
the scale of the conflict with \cite{Laney12} (and many similar values in the
literature) because we ought to have corrected our values for the systematic
distance over-estimates implied by the upper panel of \figref{fig:dist_fac}.
When this is done (using the red curve) we obtain $M_H=-1.21$ and
$M_K=-1.32$; since we have moved the stars nearer, we conclude that they are
less luminous.

\item The study of \cite{Laney12} involved obtaining new $J,H,K$ photometry
for their Hipparcos stars because the 2MASS photometry of Hipparcos red clump
stars, which have bright apparent magnitudes, is affected by saturation,
which makes them appear fainter than they really are. Unfortunately, only
four of our stars were measured by Laney et al. For these stars Laney et al.\
obtained $J$ magnitudes brighter than the 2MASS values by amounts in the
range $(-0.006,0.457)$, but their $H$ and $K$ values are not clearly brighter
than the 2MASS values, which suggests that saturation in 2MASS is mainly
confined to the $J$ band. Interestingly, the Bayesian algorithm assigns an
anomalous extinction ($A_V=0.633$) to the star (Hipp 32222) that shows by far
the strongest saturation effects, presumably because a high extinction can
explain the unexpectedly faint $J$ magnitude given the
spectroscopically determined $\Teff$. From this rather fragmentary evidence
we infer that the effects of saturation on the 2MASS magnitudes might cause
us to make the nearest clump stars under-luminous by $\sim0.1\,$mag. The
triangles in \figref{fig:dist_fac} suggest on the contrary that we have
found these stars to be over-luminous by $\sim0.4\,$mag.

\begin{figure*}
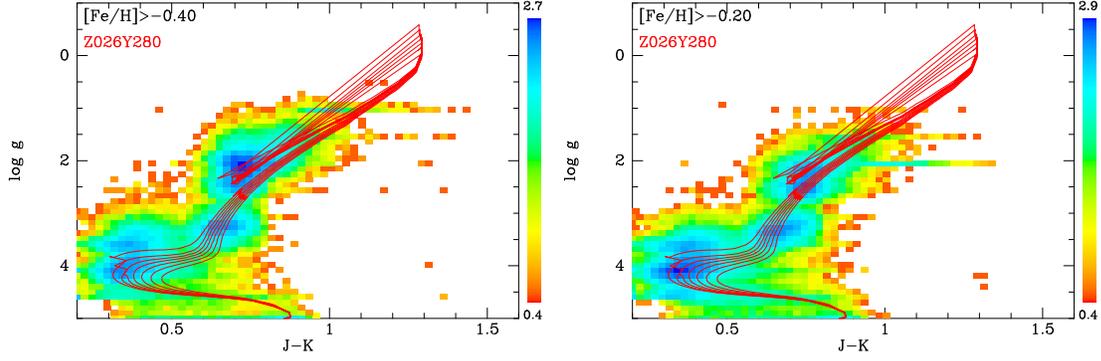

\centerline{\epsfig{file=pjm/JmKlogg1.ps,width=.4\hsize}\quad
\epsfig{file=pjm/JmKlogg2.ps,width=.4\hsize}}
\caption{Density of stars on a logarithmic scale for two metallicity ranges
in the ($J-K,\log g$)
plane together with Padua isochrones for a metal-rich populations. The left
panel is for $(-0.4<\feh<-0.2)$ and the right panel is for $(-0.2<\feh<0)$.
}\label{fig:JmKg}
\end{figure*}

\item Are the red clump stars in our sample correctly identified? \figref{fig:JmKg}
shows the density of stars in the $(J-K,\log g)$ plane for two metallicity
ranges. In both panels peaks in density are apparent near the theoretical
locations of core helium-burning stars, These peaks are captured by our
selection criteria $1.7<\log g<2.4$ and $0.55\le J-K\le0.8$. The core
helium-burning model star that sits at the centre of the red circle has
$\Teff=4485$, $\log g=2.37$ and $M_K=-1.60$, in agreement with the
empirical data of \cite{Laney12}. 
\end{itemize}

This discussion explains why our raw distances imply absolute magnitudes for
clump stars that differ little from the empirical value of Laney et al., and
why these distances are only slightly larger than the Hipparcos parallaxes
imply.  The puzzle remains that the SBA kinematic analysis points to our
distances being too large. For the SBA analysis to be correct, we would
require \emph{both} that the stellar models were too luminous \emph{and} the
Hipparcos stars to be misleading, perhaps because they are nearby and
therefore anomalously young and have atypical chemistry. Consequently,  we
set the SBA correction factors aside for the moment but in a companion paper
\citep{Binneyetal13} we will return to this issue in the context of dynamical
Galaxy models.

Table~\ref{tab:gen_ratios} shows that $1/\ex{\varpi}$ is always the shortest
of our distance measures, and given the suggestion from the SBA analysis that
even this measure might be too long, we do not present an SBA analysis of
distances based on $\ex{s}$ or $s_{\ex{\mu}}$. However, such analyses do confirm
that these measures over-estimate distances to all classes of star by even
larger factors than $1/\ex{\varpi}$ does, so there is no case to be made for
using them. 

\begin{figure*}
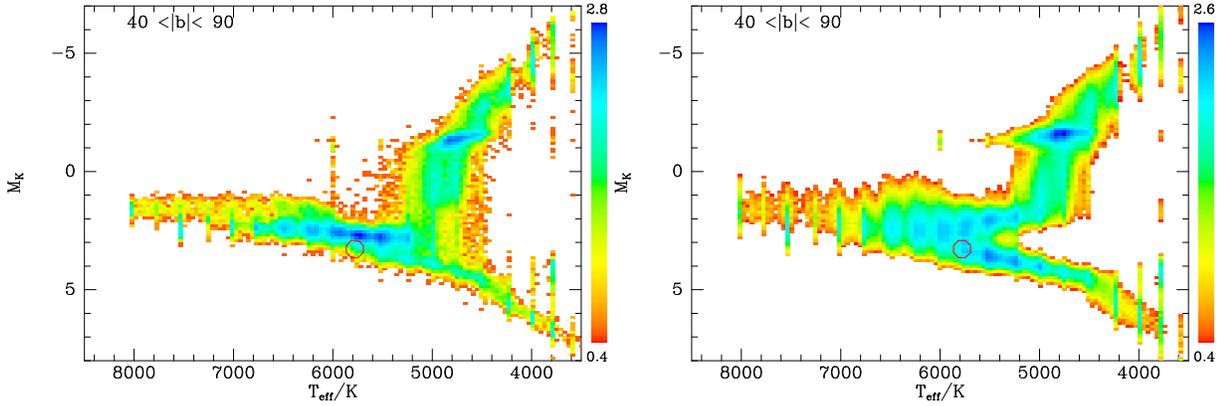

\centerline{\epsfig{file=pjm/CMD90.ps,width=.45\hsize}\ 
\epsfig{file=pjm/CMD90pdf.ps,width=.45\hsize}}
 \caption{Effective temperature absolute-magnitude diagrams for $|b|>40^\circ$ either using for each
star the absolute magnitude implied by the distance estimator $1/\ex{\varpi}$
(left) or spreading each star out in absolute magnitude according to the
multi-Gaussian representation of its pdf in distance modulus (right). The
density scale is essentially logarithm: the quantity plotted is
$\log_{10}(1+n)$, where $n$ is the number of stars in a cell. The red octagon
is centred on the location of the Sun $(\Teff,H)=(5780,3.28)$.} \label{fig:CMD}
\end{figure*}

\subsection{Effective temperature absolute magnitude diagrams}

\figref{fig:CMD} shows effective temperature absolute-magnitude diagrams for high-latitude
($|b|>40^\circ$) stars created either (a) using $\ex{\varpi}$ to assign a
single distance to each stars (left panel) or (b) spreading each star in
$M_K$ according to the multi-Gaussian fit to its pdf in distance modulus. 
The red octagon centred on $(\Teff,M_K)=(5780,3.28)$ shows the location of the
Sun in the effective temperature absolute magnitude diagram.

The red clump is prominent in both panels but the horizontal branch extends
further to the blue when the pdfs are used as a consequence of eliminating
the messy scatter of stars in the left panel between the horizontal branch
and the main sequence.  Using the pdfs similarly eliminates the unphysical
scatter of stars inside the turn-off curve. In both diagrams vertical stripes
are evident, especially at the coolest temperatures: these are a legacy of
the use by the pipeline of the DEGAS decision-tree routine to identify
template spectra \citep{DR4}. This artifact is enhanced because we
have smeared stars in $M_K$ but not in $\Teff$, as we should have done for
consistency. 

\begin{figure*}
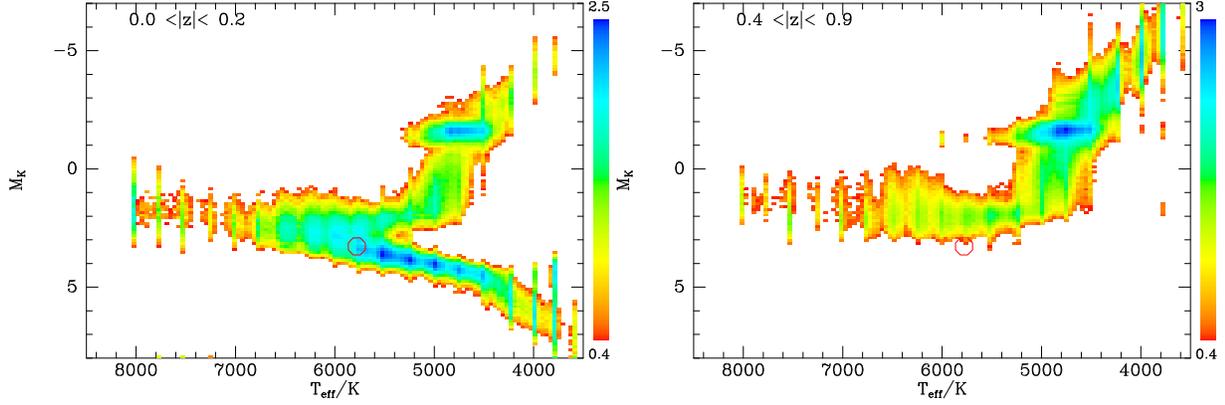

\centerline{\epsfig{file=pjm/CMDz0pdf,width=.45\hsize}\
\epsfig{file=pjm/CMDz4pdf,width=.45\hsize}}
\caption{Effective temperature absolute-magnitude diagrams for two slices in $|z|$ constructed using
the multi-Gaussian representations of stars' pdfs and using the same density
scale as in \figref{fig:CMD}.}
\label{fig:CMDz}
\end{figure*}

\figref{fig:CMDz} shows effective temperature absolute-magnitude diagrams for two slices through the
Galaxy: $|z|<0.2\kpc$ or $0.4<|z|/\!\kpc<0.9$. For these plots we used the
multi-Gaussian representations of pdfs to spread stars in distance modulus
and thus in $z$. At $|z|<0.2\kpc$ the main
sequence, subgiant and giant branches show up nicely, and the red clump is
extremely sharp.  More than $0.4\kpc$ away from the plane the lower main
sequence has disappeared and giant branch becomes more strongly populated
because the volume surveyed is much larger.

{\tabcolsep=4pt
\begin{table*}
\caption{Analysis of cluster stars. $\sigma_{\rm cl}$ is the assumed cluster
radius and $\log\tau$ gives the literature age while $\overline{\log\tau}$ is
the mean logarithm of the inferred ages of cluster stars (in years) and
$s_{\rm cl}$ is the literature value of the cluster distance (in pc).
$N_{\rm g}$ and $N_{\rm d}$ are the numbers of giants and dwarfs in the
sample and $\overline{s}_{\rm g}$ and $\overline{s}_{\rm d}$ are the mean
distances to giants and dwarfs inferred from values of $1/\ex{\varpi}$.
$\overline{s}_{\rm all}$ is the mean distance of all cluster stars and is
given also when extinction is neglected ($\overline{s}_{\rm all,nE}$ and when
a strong, cluster-specific age prior is used ($\overline{s}_{\rm all,\tau}$).
 }\label{tab:cluster_s}
\begin{center}
\begin{tabular}{lcccccccccccc}
Cluster&$\sigma_{\rm cl}$&$E(B-V)$&$\log\tau$&$\overline{\log\tau}$&$s_{\rm cl}$&
$N_{\rm g}$&$\overline{s}_{\rm g}/s_{\rm cl}$&$N_{\rm d}$&
$\overline{s}_{\rm d}/s_{\rm cl}$&$\overline{s}_{\rm all}/s_{\rm cl}$&
$\overline{s}_{\rm all,nE}/s_{\rm cl}$&$\overline{s}_{\rm all,age}/s_{\rm cl}$\\
\hline
   Blanco 1 &  5.5 &  0.01 &  7.80 &  9.59 &  269 &   4 & 1.61 &  23 & 1.07 & 1.15& 1.13& 0.87\\
   NGC 2422 &  3.6 &  0.07 &  7.86 &  8.82 &  490 &   0 & $-$ &  13 & 1.09 & 1.09& 1.08& 0.85\\
  Alessi 34 & 15.4 &  0.18 &  7.89 &  9.58 & 1100 &  24 & 1.20 &   0 & $-$ & 1.20& 1.82& 3.71\\
    ASCC 69 & 14.0 &  0.17 &  7.91 &  9.51 & 1000 &  30 & 1.63 &   2 & 0.84 & 1.58& 2.11& 4.87\\
   NGC 6405 &  2.8 &  0.14 &  7.97 &  8.89 &  487 &   0 & $-$ &  12 & 0.94 & 0.94& 0.90& 0.71\\
 Melotte 22 (Pleiades) &  4.6 &  0.03 &  8.13 &  9.39 &  133 &   2 & 1.11 &  35 & 1.09 & 1.09& 1.11& 0.93\\
   NGC 3532 &  7.1 &  0.04 &  8.49 &  8.95 &  486 &   1 & 1.71 &  17 & 1.23 & 1.26& 1.24& 0.97\\
   NGC 2477 (M93)&  5.7 &  0.24 &  8.78 &  9.29 & 1300 &  45 & 0.91 &   3 & 1.02 & 0.92& 1.13& 1.27\\
     Hyades &  5.7 &  0.01 &  8.80 &  9.70 &   46 &   0 & $-$ &  31 & 1.08 & 1.08& 1.08& 1.00\\
   NGC 2632 (Praesepe, M44)&  3.8 &  0.01 &  8.86 &  9.48 &  187 &   0 & $-$ &  34 & 1.14 & 1.14& 1.12& 1.03\\
   NGC 2423 &  2.7 &  0.10 &  8.87 &  8.96 &  766 &   3 & 1.36 &  17 & 0.99 & 1.05& 0.99& 1.17\\
    IC 4651 &  2.6 &  0.12 &  9.06 &  9.30 &  888 &   7 & 1.03 &   5 & 0.68 & 0.87& 0.87& 1.00\\
   NGC 2682 (M67) &  6.6 &  0.06 &  9.41 &  9.74 &  908 &  32 & 0.74 &  12 & 0.65 & 0.72& 0.74& 0.78\\
\hline
\end{tabular}
\end{center}
\end{table*}
}
\section{Cluster stars}\label{sec:clusters}

By searching for stars that have suitable sky coordinates and line-of-sight
velocities that agree with a cluster convergence point, we have identified
RAVE stars in 15 open clusters. NGC 3680 has just one RAVE
star so we cannot analyse its statistics.  Table~\ref{tab:cluster_s}
lists the remaining clusters with RAVE stars in order of increasing age,
giving for each cluster the values of several quantities from the literature.
The values given are taken from \cite{Dias02} with the exception of the Hyades,
where we used \cite{Perryman98}.

Columns 7 and 8 give the number of giants in our sample and the ratio of
their mean value of $1/\ex{\varpi}$ to the distance listed in
Table~\ref{tab:cluster_s}.  Columns 9 and 10 give the same data for dwarfs,
and column 11 gives the overall mean of $1/\ex{\varpi}$ for cluster stars
divided by the literature distance.  A tendency for the giants to
over-estimate distances is evident, particularly in the younger clusters such
as Alessi 34 and ASCC 69. The distances inferred for dwarfs are generally in
good agreement with the literature values, but significant under-estimates
are evident in the cases of the oldest clusters, IC 1651 and NGC 2682 (M67).
The penultimate column gives the mean value of $1/\ex{\varpi}$ divided by the
literature distance when extinction is assumed to be zero.  Setting $A_V=0$
shortens distances to dwarfs and lengthens those to giants and for a few
clusters the results with no dust are markedly worse but neglecting dust has
little impact on most clusters. 

Column 5 gives the the mean inferred value of the logarithm of age (in years)
and comparing these values with the literature values in column 4 we see
little sign of correlation with the result that stars in younger clusters are
being presumed much older than they really are. This phenomenon reflects the
fact that dating an isolated star is enormously harder than dating a cluster
of coeval stars. Clearly poor ages will bias the recovered distances so in
the last column of Table~\ref{tab:cluster_s} we give the mean values of
$1/\ex{\varpi}$ divided by the literature distance when distances are
determined under the strong age prior
 \[
P(\tau)\propto\exp\left[-\log_{10}^2(\tau/\tau_{\rm cl})/2(0.1)^2\right],
\]
 This cluster-specific age prior improves the accuracy of mean distances to
stars in clusters older than $100\Myr$, but has an unfortunate effect on the
distances to stars in younger clusters. 

\begin{figure*}
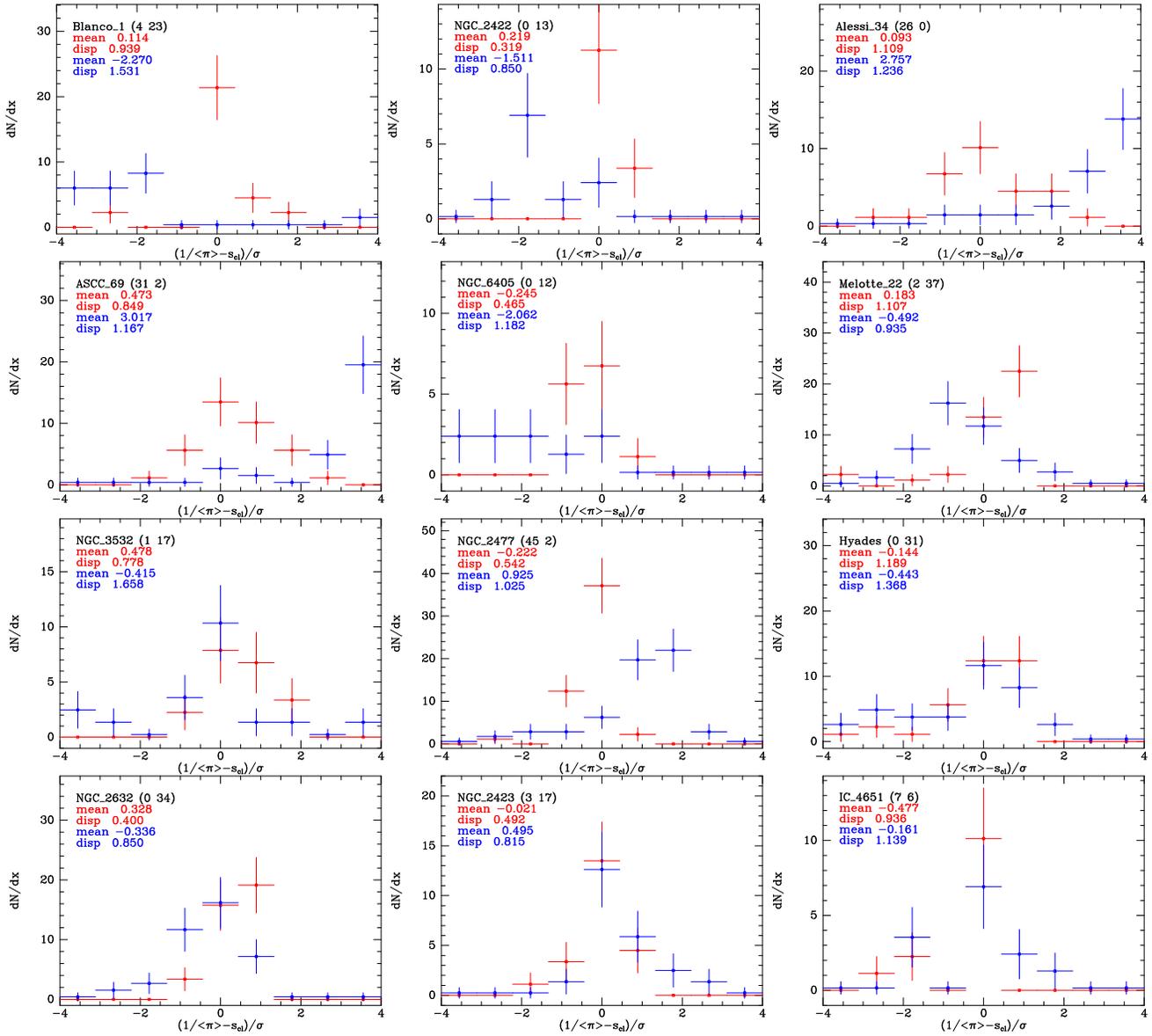

\centerline{
\epsfig{file=pjm/blanco_1_s.ps,width=.32\hsize}
\epsfig{file=pjm/NGC_2422_s.ps,width=.32\hsize}
\epsfig{file=pjm/Alessi_34_s.ps,width=.32\hsize}
}
\centerline{
\epsfig{file=pjm/ASCC_69_s.ps,width=.32\hsize}
\epsfig{file=pjm/NGC_6405_s.ps,width=.32\hsize}
\epsfig{file=pjm/melotte_22_s.ps,width=.32\hsize}}
\centerline{
\epsfig{file=pjm/NGC_3532_s.ps,width=.32\hsize}
\epsfig{file=pjm/NGC_2477_s.ps,width=.32\hsize}
\epsfig{file=pjm/hyades_s.ps,width=.32\hsize}}
\centerline{
\epsfig{file=pjm/NGC_2632_s.ps,width=.32\hsize}
\epsfig{file=pjm/NGC_2423_s.ps,width=.32\hsize}
\epsfig{file=pjm/IC_4651_s.ps,width=.32\hsize}}
 \caption{Histograms (in red) of the distances $1/\ex{\varpi}$ to stars in
individual clusters (see \figref{fig:m67} for NGC 2682). The blue points show
results obtained when the prior on the age is a Gaussian in $\log\tau$ with
dispersion $0.1\Gyr$ and centred on the literature value given in
Table~\ref{tab:cluster_s}.  The clusters are ordered from top left to bottom
right by age. The numbers in brackets after the cluster name give the number
of giants and dwarfs contributing to the plot. The normalising dispersion $\sigma$ is the
quadrature sum of the error on the distances and the size of the cluster
listed in Table~\ref{tab:cluster_s}. The blue points have been moved up
slightly for clarity.}\label{fig:cluster_his_s}
\end{figure*}

\begin{figure}
\centerline{\epsfig{file=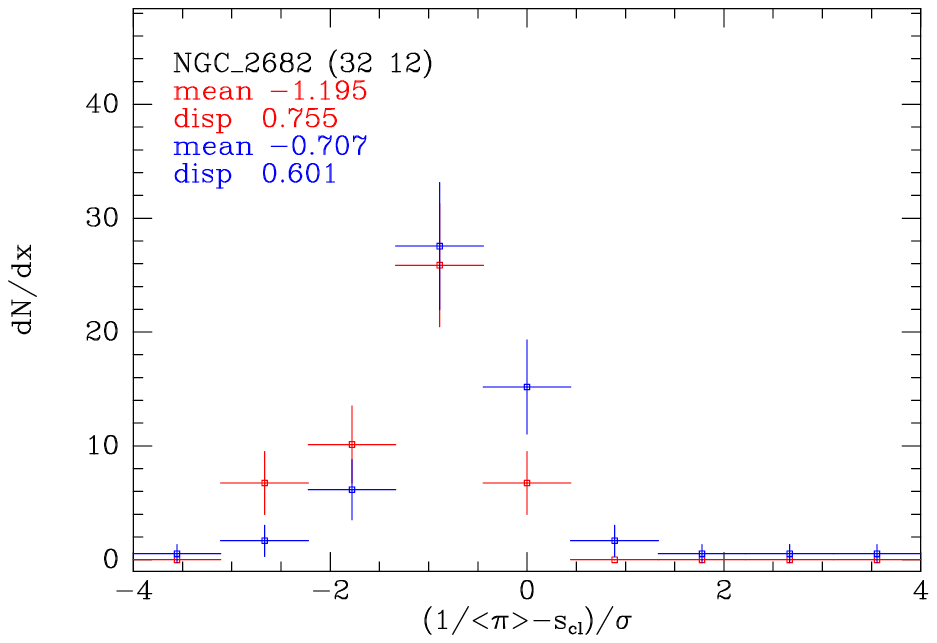,width=.8\hsize}}
\centerline{\epsfig{file=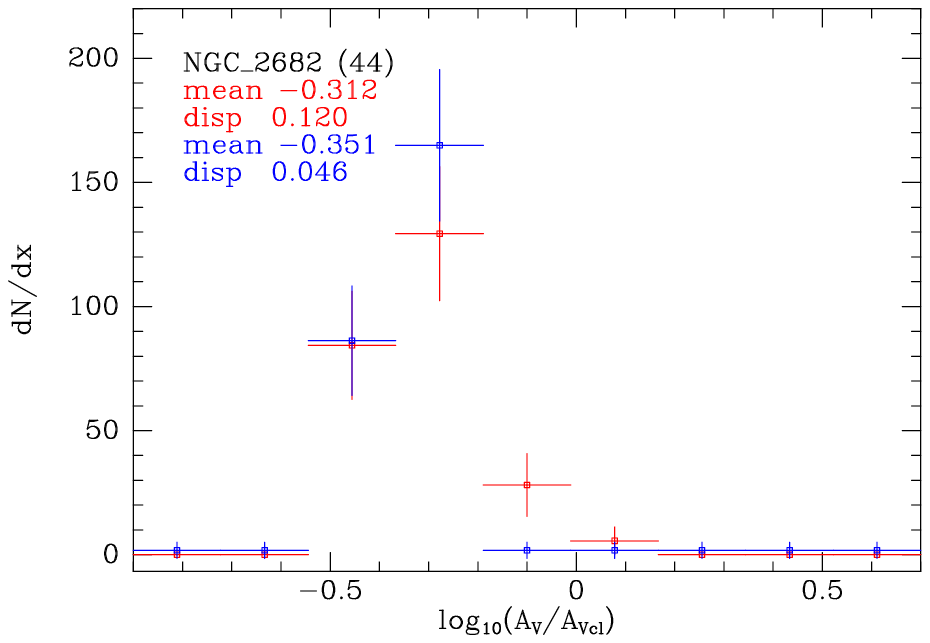,width=.8\hsize}}
\caption{Upper panel: histogram of distances to stars in NGC 2682 (M87). Red
points are obtained with the standard age prior, blue points with prior that
specifies the literature age of the cluster. Lower panel: histogram of ratios
of extinctions of stars to the cluster's literature extinction. The red
points are the Bayesian extinctions and the blue points the priors from the
Schlegel et al.\ map.}\label{fig:m67}
\end{figure}

\figref{fig:cluster_his_s} shows histograms of distances to stars in 12 of
the 13 clusters listed in Table~\ref{tab:cluster_s}; the red histograms are
for our standard distances and the blue histograms are for distances obtained
under the strong cluster-specific prior. The numbers in brackets after the
cluster names in the top left corner of each panel give the number of giants
and dwarfs in that cluster. The top panel of \figref{fig:m67} shows the
corresponding plot for NGC 2682 (M67).  We see that the strong age prior
shortens distances to dwarfs and lengthens those to giants in a way that is
moderate and beneficial in clusters as old as the Melotte 22 (Pleiades) but
unhelpful in younger clusters. The red histograms are generally quite
satisfactory.

\begin{figure}
\centerline{\epsfig{file=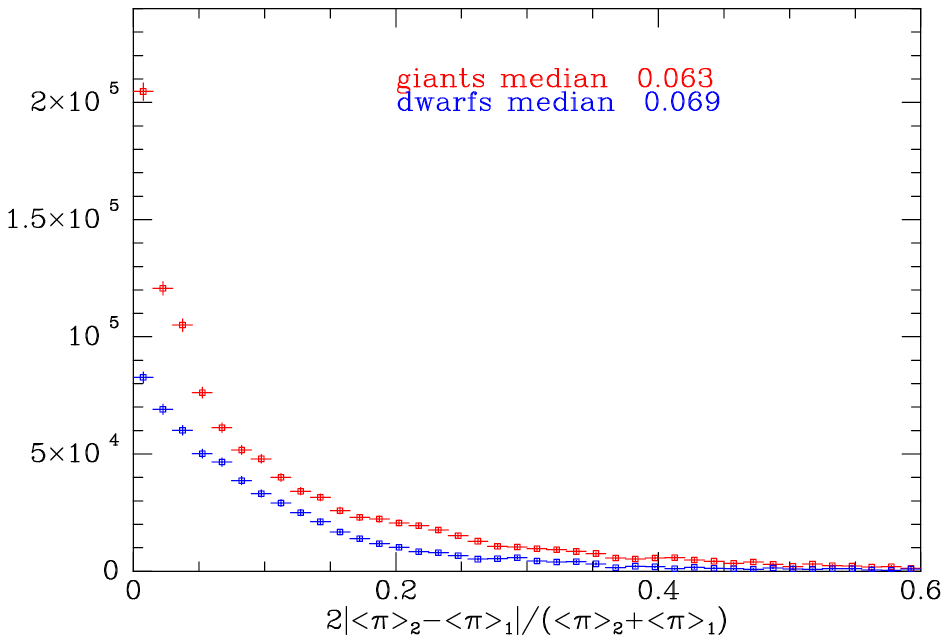,width=.8\hsize}}
\centerline{\epsfig{file=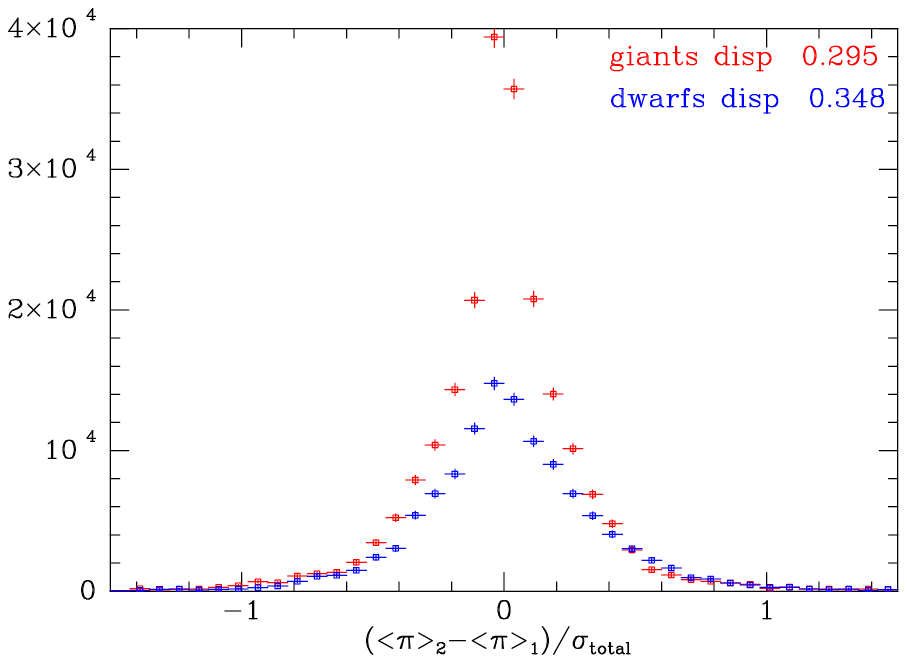,width=.8\hsize}}
\caption{Discrepancies between different measurements of the distances of the
same star.}\label{fig:pair_s}
\end{figure}

\section{Repeat observations}\label{sec:repeat}

We have more than one spectrum for $12\,012$ stars and can form $8\,526$
independent pairs of measurements for the same dwarf star and $11\,868$
independent pairs of measurements for the same giant star.
\figref{fig:pair_s} shows histograms of the discrepancies between these
measurements when normalised in two ways. In the upper panel the difference
in $\ex{\varpi}$ is divided by the mean parallax, while in the lower panel it
is divided by the quadrature-sum of the uncertainties of the measurements.
The median fractional parallax discrepancy is $0.063$ for giants and $0.069$
for dwarfs -- it is easy to show that these values apply also to the
discrepancies in distances $1/\ex{\varpi}$. The dispersions of the parallax
discrepancies normalised by the formal uncertainties is $0.295$ for giants
and $0.348$
for dwarfs. That these numbers are significantly smaller than unity
emphasises that much of the error is external and does not derive from noise
in the spectrum.

\begin{figure}
\centerline{\epsfig{file=pair_Av.ps,width=.8\hsize}}
\caption{Discrepancies between different measurements of the extinctions to the
same star.}\label{fig:pair_Av}
\end{figure}

\begin{figure*}
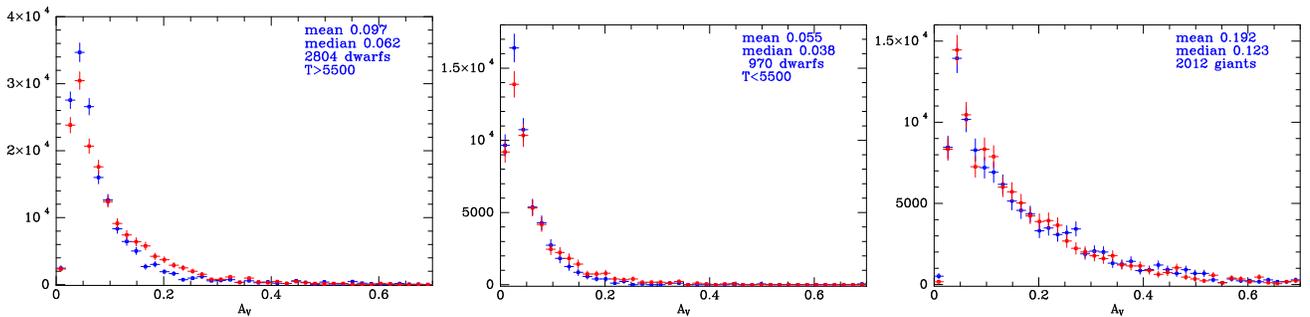

\centerline{\epsfig{file=pjm/hot_AvH.ps,width=.32\hsize}
\epsfig{file=pjm/cool_AvH.ps,width=.32\hsize}
\epsfig{file=pjm/giant_AvH.ps,width=.32\hsize}}
\caption{Blue points show the distributions of $\widetilde A_V$ for Hipparcos
stars: hot dwarfs
(left), cool dwarfs (centre) and giants (right). Red points show the distribution of the values of the prior
extinction at the predicted locations of the stars.}\label{fig:Av}
\end{figure*}

\begin{figure}
\centerline{\epsfig{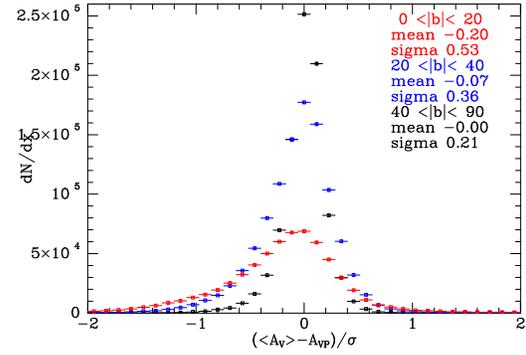}}
 \caption{Histogram of the offsets between the estimated visual extinction
$\widetilde A_V$ to stars in the complete RAVE sample and the extinction in
the dust model used as a prior to the location
$(l,b,1/\ex{\varpi})$.}\label{fig:AVP}
\end{figure}

\begin{figure*}
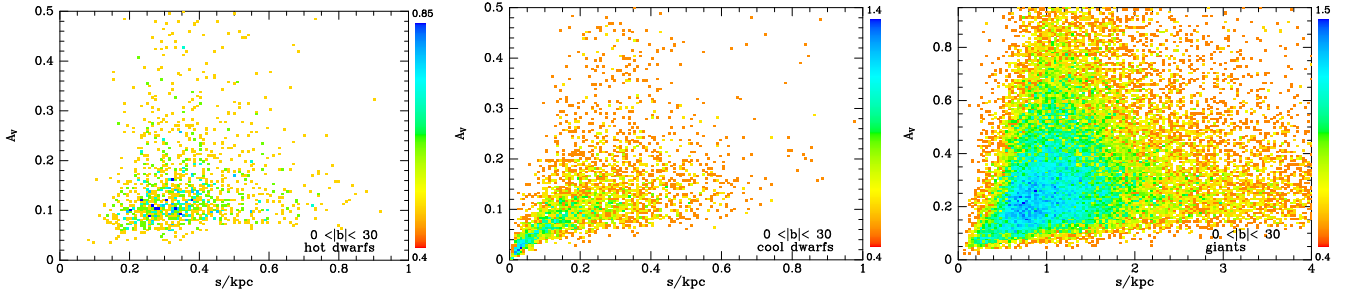

\centerline{\epsfig{file=pjm/sAvhd.ps,width=.32\hsize}\quad
\epsfig{file=pjm/sAvcd.ps,width=.32\hsize}\quad
\epsfig{file=pjm/sAvg.ps,width=.32\hsize}}
 \caption{Density of stars in the distance versus extinction plane
for hot dwarfs, cool dwarfs and giants in the range of Galactic latitudes
$|b|<30^\circ$. The density scale is as in \figref{fig:CMD}.}\label{fig:sAv}
\end{figure*}

\begin{figure*}
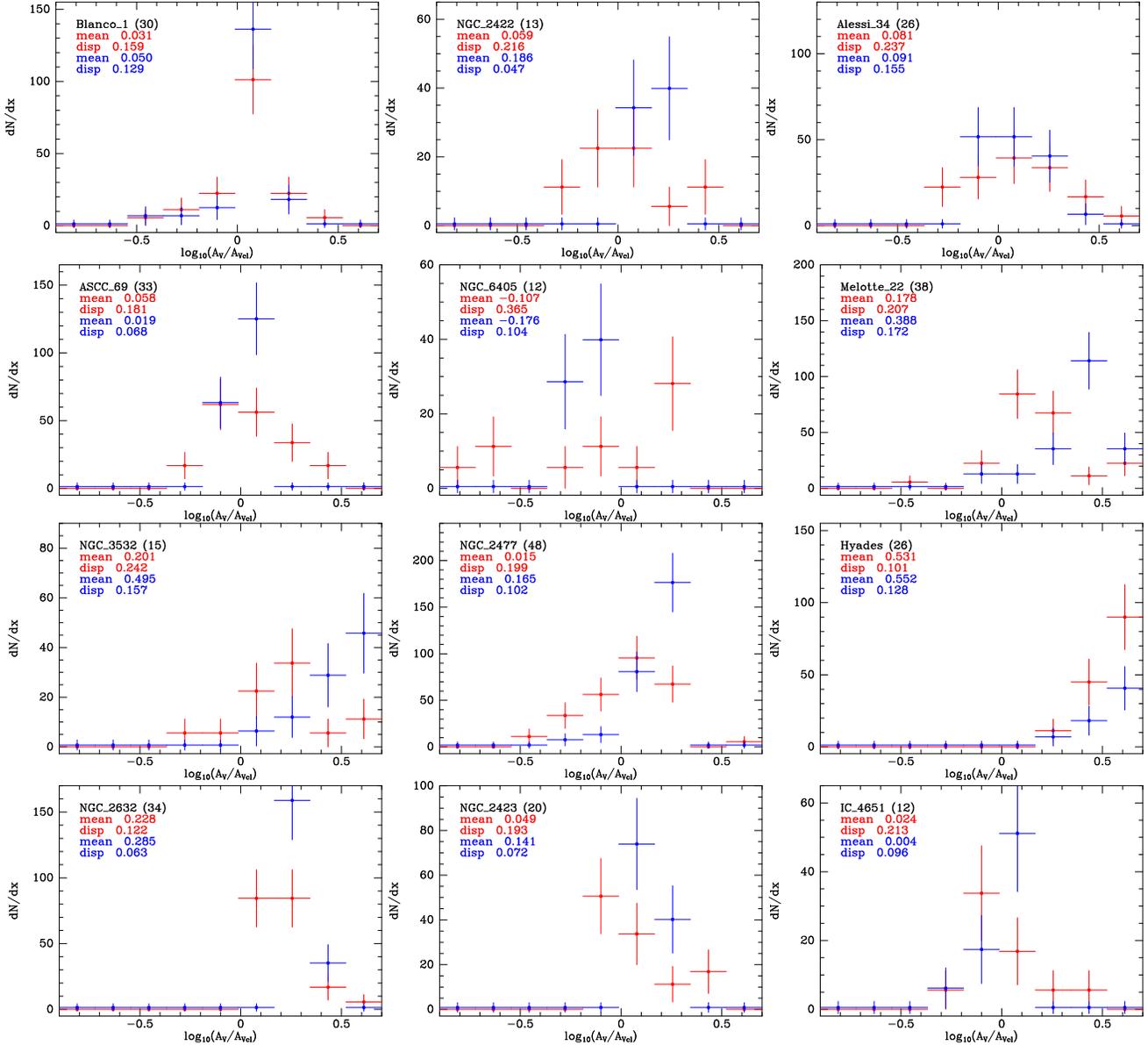

\centerline{
\epsfig{file=pjm/blanco_1_Av.ps,width=.32\hsize}
\epsfig{file=pjm/NGC_2422_Av.ps,width=.32\hsize}
\epsfig{file=pjm/Alessi_34_Av.ps,width=.32\hsize}
}
\centerline{
\epsfig{file=pjm/ASCC_69_Av.ps,width=.32\hsize}
\epsfig{file=pjm/NGC_6405_Av.ps,width=.32\hsize}
\epsfig{file=pjm/melotte_22_Av.ps,width=.32\hsize}}
\centerline{
\epsfig{file=pjm/NGC_3532_Av.ps,width=.32\hsize}
\epsfig{file=pjm/NGC_2477_Av.ps,width=.32\hsize}
\epsfig{file=pjm/hyades_Av.ps,width=.32\hsize}}
\centerline{
\epsfig{file=pjm/NGC_2632_Av.ps,width=.32\hsize}
\epsfig{file=pjm/NGC_2423_Av.ps,width=.32\hsize}
\epsfig{file=pjm/IC_4651_Av.ps,width=.32\hsize}}
 \caption{As \figref{fig:cluster_his_s} but showing in red of the offsets of $\log_{10}\widetilde A_V$
from the cluster's literature value of $\log_{10}A_V$ (see \figref{fig:m67}
for NGC 2682).
The blue points show
the amounts by which the prior extinction at $1/\ex{\varpi}$ differs from the
literature value. The blue points have been moved up slightly for
clarity.}\label{fig:cluster_his_Av}
\end{figure*}

\section{Estimated extinctions}\label{sec:Av}

As with distances, the Bayesian algorithm determines a probability
distribution for possible extinctions to each star, and one has to consider
how best to reduce this distribution to a single value for the extinction.
For the reasons given in Section~\ref{sec:method} the code marginalises over
extinctions by integrating with respect to $a\equiv\ln(A_V)$ rather than
integrating with respect to $A_V$ directly. Consequently a natural quantity
to output is $\ex{a}$, and we use $\widetilde A_V\equiv\e^{\ex{a}}$ as our
estimator of the extinction. $\widetilde A_V$ places less weight on high
extinctions than does $\ex{A_V}$.

\figref{fig:pair_Av} shows that different spectra yield the same value for
$\widetilde A_V$ to high precision: the dispersion in the differences divided
by the quadrature sum of the uncertainties is only $0.117$ for giants and
$0.097$ for dwarfs. This result is to be expected because $\widetilde A_V$
depends strongly on the photometry, and we only change the spectrum between
determinations of $\widetilde A_V$.

\figref{fig:Av} shows in red the distribution of extinctions to Hipparcos
stars; the blue points show the distribution of the prior values of the
extinction to the final locations $1/\ex{\varpi}$ of the stars. Since the
red and blue points follow very similar distributions, on average our
recovered extinctions coincide well with our priors. This finding could
indicate either that our priors are accurate guesses of the actual
extinction, or that the extinction to an individual star cannot be determined
from the data we have. We know that the data {\it are\/} adequate because
when we took the priors from the smooth model (\ref{eq:dmodel}) normalised in
an average sense by the {Schlegel} et al.\ reddenings, the recovered values of 
$\widetilde A_V$ were systematically smaller than the prior values. Thus the
data suffice to shift the recovered values away from a poor prior.
Presumably the Hipparcos stars lie in directions of anomalously low
extinction, an effect that is captured when the extinction is estimated to be
the fraction of the measured extinction to infinity that is expected to lie
within distance $s$.

For hot dwarfs most values of
$\widetilde A_V$ lie in $(0.1,0.25)$ [so $\widetilde A_J$ lies in
$(0.03,0.07)$], while a significant fraction of cool dwarfs have $\widetilde
A_V<0.1$ as we would expect given that some of these stars are quite close.
The distribution of values of $\widetilde A_V$ for giants peaks around $0.2$
but has a long tail extending out to $\sim0.6$ as we expect for stars that
can be quite distant. 

\figref{fig:AVP} shows histograms of the differences between our estimated
extinctions $\widetilde A_V$ to stars in the complete sample and the value of
the prior on extinction to the star's proposed location. The red, blue and
black histograms are for stars that lie in three ranges of Galactic latitude
$b$.  The means of all the two highest-latitude histograms are satisfyingly
close to zero. The mean of the histogram for $|b|<20\,$deg is negative
($-0.2\sigma$) implying that the dust model slightly over-estimates
extinctions to low-latitude stars.

\figref{fig:sAv} shows the relationship between extinction and distance for
hot dwarfs ($\Teff>5500\K$), cool dwarfs and giants ($\log g<3.5$) in the
full RAVE sample. In addition to showing the extent of the relation between
distance and extinction, these plots show how the three classes of star are
distributed in distance. The ridge line through the distribution of giants
has a slope $\simeq0.19\,$mag/kpc, while that through the distribution of
cool dwarfs has a slope $\simeq0.78\,$mag/kpc. For comparison, the
traditional relation for paths near the mid-plane is $A_V\simeq1.6 s/$kpc
\citep[e.g.][]{BinneyM}.  Since most of our sight lines move away from the
mid-plane, they naturally have lower values of extinction per unit length.
Moreover, our samples are subject to the already-noted observational bias
against stars high extinctions, and this bias particularly concentrates the
giants at high latitudes, where extinction per unit distance is low.

The red points in \figref{fig:cluster_his_Av} show for each cluster the
distribution of $\log_{10}(\widetilde A_V/A_{V{\rm cl}})$, where $A_{V{\rm
cl}}$ is 3.1 times the cluster's literature value of $E(B-V)$.  The blue
points show the corresponding distributions of the values obtained by
replacing $\widetilde A_V$ by the prior extinction $A_{V{\rm prior}}$ at
$1/\ex{\varpi}$. For all clusters the red and blue points have similar
distributions, which suggests that the priors are reasonable. In light of
this result, it is striking (a) how broad the distributions are, and (b) that
in four clusters (Melotte 22, Hyades, NGC 2632 and NGC 2682) the literature extinction
lies off one wing or the other of the distribution. These findings call into
question the very concept of a cluster-wide characteristic extinction, and
suggest that if one must choose a single characteristic extinction, the
literature value may be a poor choice.

\section{Conclusions}\label{sec:discuss}

We have extended the Bayesian approach to distance determination of
\cite{BurnettB} to allow for extinction and reddening and to deliver pdfs in
distance modulus in addition to expectation values of three distance
measures, distance $s$, distance modulus $\mu$ and parallax $\varpi$. 

We have fitted each star's pdf in distance modulus with a sum of up to three
Gaussians. A single Gaussian provides a good fit to about 45 per cent of the
pdfs, two Gaussians provide a good fit to most of the remaining pdfs, so just
5 per cent of the pdfs require three Gaussians for a good fit. When these
Gaussian decompositions are used to make Hess diagrams by splitting each
star's contribution to the density into one, two or three parts at the
luminosity associated with the centre of each Gaussian component, the diagram
becomes significantly sharper as the man-sequence turnoff and the horizontal
branch emerge clearly. This phenomenon indicates that multi-modal pdfs are
associated with stars that could be upper main-sequence stars or blue
horizontal-branch stars, or could be lower main-sequence stars or subgiants.

For every class of star examined, we find that
$\ex{s}>s_{\ex{\mu}}>1/\ex{\varpi}$, a phenomenon that arises because these
distance measures weight differently the possibilities that a given star is
far or near. The differences between these distance measures are least for
hot dwarfs $\Teff>5500\K$ and red-clump stars, and greatest for very cool
dwarfs ($\Teff<5200\K$) high-gravity giants ($\log g>2.4$) because hot dwarfs
and red-clump stars have quite narrow pdfs in distance while the dwarf/giant
ambiguity causes cool dwarfs and high-gravity giants to have broad pdfs in
distance.

The RAVE survey encompasses $\sim5000$ Hipparcos stars.  Histograms of the
difference between our values of $\ex{\varpi}$ and the Hipparcos parallaxes
normalised by the quadrature sum of our errors and the Hipparcos errors come
close to the ideal of a unit Gaussian of zero mean in the cases of warm
dwarfs ($\Teff>5500\K$) and giants ($\logg<3.5$), so not only are our
parallax estimates fairly reliable, but our error estimates are reasonable.
The situation regarding the smaller sample of cool dwarfs is unsatisfactory.
The majority of these stars require multi-Gaussian fits to their pdfs. When
a Hipparcos parallax is available, it agrees within the errors with one of
the Gaussians as one would wish. But the single Gaussians fitted to a
minority of cool dwarfs yield parallaxes that are significantly larger than
the Hipparcos parallaxes. Thus our ability to determine distances to cool
dwarfs is rather limited.

For giants our parallaxes are competitive with those of Hipparcos, but for
cool dwarfs errors on Hipparcos parallaxes are smaller than the errors on
ours by a factor $\sim3$.

The good agreement between our parallaxes and the Hipparcos parallaxes,
suggests that $1/\ex{\varpi}$ is our most reliable estimator of distance, a
conclusion we were able to confirm subsequently. Hence we have concentrated
on assessing the accuracy of the distance estimator $1/\ex{\varpi}$.

The Hipparcos stars in the RAVE survey reveal (\figref{fig:dist_fac}) a
tendency for our distances to the hottest dwarfs to be $\sim15\%$ too small,
while our distances to dwarfs with $\Teff\sim5000\K$ are too large by about
the same amount. Our distances to the coolest dwarfs are 20--30\% too small.
The Hipparcos stars reveal that our distances to giants are too large by a
factor that increases smoothly with decreasing $\log g$ from unity at $\log
g=3.5$ to $\sim1.2$ at the lowest gravities. This phenomenon may reflect our
use of stellar parameters obtained under the assumption of LTE. However, it
should be noted that \cite{DR4} excise the cores of strong lines, where
non-LTE effects will be most prominent.

The values of the kinematic corrections obtained by the method of \cite{SBA}
for all the giants and dwarfs in the RAVE sample confirm the results from the
Hipparcos stars: $1/\ex{\varpi}$ is a more reliable distance estimator for
cool stars than $\ex{s}$ and for dwarfs the ratio of $1/\ex{\varpi}$ to the
true distance increases with decreasing $\Teff$ except below
$\Teff\sim4500\K$, where it drops abruptly. For dwarfs the SBA kinematic
indicators agree moderately with each other and suggest that our distances
tend to be too short by an amount that decreases with $\Teff$ from $\gta20\%$
at the hot end to perfection at $\Teff\simeq5000\K$. The shape of the plot of
the ratios of our distance to true distance agrees perfectly with the
Hipparcos results, but there is a small vertical offset between the curves.

For giants $1/\ex{\varpi}$ has a tendency to be too large, by an amount that
emerges equally from the Hipparcos results and the SBA kinematic corrector
$f_W$. The ratio of our distance to the true distance increases with
decreasing $\log g$ from $\sim1.05$ at the high-gravity end to $\sim1.2$
at the low-gravity end. Unfortunately, the Hipparcos results are of course
confined to $s\lta0.15\kpc$ and the SBA analysis proves sensitive to the
upper limit on the distances of stars we use in the analysis. Moreover, for
stars with $s\gta2\kpc$ the two SBA factors disagree with each other.
Therefor it is hard to assess the accuracy of our distances to stars at
$s>2\kpc$, which tend to be luminous low-gravity giants. However, the
indications are that we are over-estimating these distances by $\gta20\%$.

We have identified red-clump stars by cuts in the $(J-K,\log g)$ plane and
find that a histogram of these stars' values of $M_K$ is narrow and peaks
$\sim0.1\,$mag fainter than the standard magnitude. The origin of this offset
is unclear. If we accept the indications from both the Hipparcos stars and
the SBA analysis that we systematically over-estimate distances to giants,
the offset is made significantly larger: $0.3\,$mag under-luminous.

We have identified 364 RAVE stars in 15 open clusters. Our standard distances
generally form a satisfyingly narrow distribution with the cluster's
literature distance almost always within one standard deviation of the
distribution's mean. There is a clear tendency for the giants in any cluster
to be assigned distances that are larger than the distances assigned to the
cluster's dwarfs. In the oldest clusters, IC 4651 and NGC 2682 (M67), the
dwarf distances are only $\sim67\%$ of the cluster distance, but in the other clusters
the dwarf distances appear about right.

The data barely constrain the ages of stars. Consequently, our standard
distances are based the assumption that stars are quite old, older than the
ages of many of the clusters we have studied. Curiously, using a prior on
ages that enforces the cluster's literature age produces a more satisfying
histogram of distances only for clusters older than Melotte 22 (the
Pleiades).

The data do contain sufficient information to place significant constraints
on the extinctions of stars -- we know this because the extinctions we first
derived were systematically lower than the priors we then employed. This
phenomenon led to improved priors and our extinctions now scatter nearly
randomly around the prior values. Since extinction varies discontinuously
from one line of sight to the next on account of the fractal nature of the
ISM, and we do not have a sample of stars with accurately determined
extinctions, it is hard to test the validity of our extinctions. Our results
for clusters indicate that different stars in the same cluster generally have
significantly different extinctions, and that the mean extinction of stars in
a given cluster often differs significantly from the cluster's literature
value.

The distances we derive from different spectra of the same star are entirely
consistent with one another and imply that noise in the spectrum contributes
less that half the uncertainty in the derived distance. 

This work could and should be significantly improved in three ways. First,
photometry in optical bands is now available for most of our stars from the
APASS survey \citep{Henden}. Use of this photometry would sharpen constraints
on some combination of $A_V$ and $\Teff$. Second, the stellar models used
here are now a few years old and should be updated and extended. Inclusion of
$\alpha$-enhanced stars with lower metallicities should improve accuracy for
stars that are far from the plane. Moreover, we could now use models for
which magnitudes in the 2MASS system have been directly computed rather than
obtained by transformation of magnitudes in the Johnson-Cousins system.
Third, stellar parameters that include corrections for non-LTE effects as
discussed by \cite{Ruchti} may yield improved distances, especially to
luminous giants.  Distances based on extended photometry and models will be
made available on the RAVE website as soon as possible.

\section*{Acknowledgements}

We thank the referee for a meticulous reading of the submitted version and
many useful suggestions for improvement.

Funding for RAVE has been provided by: the Australian Astronomical Observatory;
the Leibniz-Institut f\"ur Astrophysik Potsdam (AIP); the Australian
National University; the Australian Research Council; the French National Research
Agency; the German Research Foundation (SPP 1177 and SFB 881); the
European Research Council (ERC-StG 240271 Galactica); the Istituto Nazionale
di Astrofisica at Padova; The Johns Hopkins University; the National Science
Foundation of the USA (AST-0908326); the W. M. Keck foundation; the Macquarie
University; the Netherlands Research School for Astronomy; the Natural
Sciences and Engineering Research Council of Canada; the Slovenian Research
Agency; the Swiss National Science Foundation; the Science \& Technology
Facilities Council of the UK; Opticon; Strasbourg Observatory; and the
Universities of Groningen, Heidelberg and Sydney. The RAVE web site is at
http://www.rave-survey.org.

\label{lastpage}

\begin{thebibliography}{}

\bibitem[Antoja et al.(2012)]{Antoja}
Anotoja T., Helmi A., Bienaym\'e O., Bland-Hawthorn J., \& the RAVE
collaboration, 2012, MNRAS, 425, L1

\bibitem[Arce \& Goodman(1999)]{ArceGoodman}
Arce H.G. \& Goodman A.A., 1999, ApJ, 512, L135

\bibitem[Aumer \& Binney(2009)]{Aumer}
Aumer M. \& {Binney} J.J., 2009, MNRAS, 397, 1286

\bibitem[Bertelli et al.(2008)]{Bertelli}
Bertelli G.,  Girardi L., Marigo P., Nasi E., 208, {A\&A}, 484, 815

\bibitem[Binney(2011)]{BinneyBangalore}
Binney J, 2011, Prama, 77, 39

\bibitem[Binney et al.(2013)]{Binneyetal13}
Binney J., \& the RAVE colaboration, 2013, to be submitted

\bibitem[Binney et al.(1997)]{BinneyGS}
Binney J.J., Gerhard O.E., Spergel D., 1997, MNRAS, 288, 365

\bibitem[Binney \& Merrifield(1998)]{BinneyM}
Binney J., Merrifield M., 1998, ``Galactic Astronomy'', Princeton University
Press, Princeton

\bibitem[Breddels et al.(2009)]{Breddels}
Breddels M.A., et al., 2010, A\&A, 511, 90

\bibitem[Burnett \& Binney(2010)]{BurnettB}
Burnett B \& Binney J., 2010, MNRAS, 407, 339

\bibitem[Burnett et al.(2011)]{Burnettetal}
Burnett B., Binney J. \& the RAVE collaboration, 2011, A\&A, 532, 113

\bibitem[Cannon(1970)]{Cannon70}
Cannon R.D., 1970, MNRAS, 150, 111

\bibitem[Carollo et al.(2009)]{Carollo}
Carollo D., Beers T.C., Chiba M., Norris J.E., Freeman K.C., Lee Y.S.,
Ivezic Z., Rockosi C.M. Yanny B., 2010, ApJ, 712, 692

\bibitem[Dehnen(1998)]{Dehnen98}
Dehnen W., 1998, AJ, 115, 2384

\bibitem[Dias et al.(2002)]{Dias02}
Dias W.S., Alessi B.S., Moitinho A., Lepine J.R.D., 2002, A\&A, 389, 871

\bibitem[Famaey et al.(2005)]{Famaey}
Famay B., Jorissen A., Luri X., Mayor M., Udry S., Dejonghe H., Turon C.,
2005, A\&A, 430, 165

\bibitem[Gillessen et al.(2009)]{Gillessen}
Gillessen S. Eisenhauer F. Trippe S., Alexander T., Genzel R., Martins F., Ott T.,
ApJ, 692, 1075

\bibitem[Haywood(2001)]{Haywood}
Haywood M., 2001, MNRAS, 325, 1365

\bibitem[Henden et al.(2012)]{Henden}
Henden A.A., Levine S.E., Terrell D., Smith T.C., Welch D., 2012, JAVSO, 40,
430

\bibitem[Juri\'c et al.(2008)]{Juric_cut}
Juri{\'c} M., Ivezi{\'c} {\v Z}, Brooks A., et al., ApJ, 673, 864

\bibitem[Koen et al.(2007)]{Koen07}
Koen C., Marang F., Kilkenny D., et al., 2007, MNRAS, 380, 1433


\bibitem[Kordopatis et al.(2013)]{DR4}
Kordopatis G., Gilmore G., Steinmetz M., Boeche C., Seabroke G.M., Siebert
A., Zwitter T.,  
de Laverny P.,  Recio-Blanco A.,
al., 2013, ApJ in press (arXiv:1309.4284)

\bibitem[Kroupa et al.(1993)]{Kroupa}
Kroupa P., Tout C.A.,  Gilmore G., 1993, MNRAS, 262, 545

\bibitem[Laney et al.(2012)]{Laney12}
Laney C.D., Joner M.D., Pietrzynski G., 2012, MNRAS, 419, 1637

\bibitem[Perryman et al.(1998)]{Perryman98}
Perryman, M.A.C., Brown A.G.A., Lebreton Y., G\'omez A., Turon C., Cayrel de
Strobel G., Mermilliod J.C., Robichon N., Kovalevsky J., Crifo F., 1998, A\&A.,
331, 81

\bibitem[Pietrzynski(2003)]{Pietrzynski}
Pietrzynski G., Gieren W., Udalski A., 2003, AJ, 125, 2494

\bibitem[Reddy(2009)]{Reddy}
Reddy B.E., 2009, in {\it Chemical Abundances in the Universe}, IAU
Symposium 265, K. Cunha, M Spite \& B. Barbuy eds, Cambridge University Press

\bibitem[Rieke \& Lebofsky(1985)]{RiekeL85}
Rieke G.H., Lebofsky R.M., 1985. ApJ, 288, 618

\bibitem[R\"oser et al.(2008)]{Roeseretal}
R\"oser S., Schilbach E., Schwan H., Kharchenko N.V., Piskunov A.E., Scholz
R.-D., 2008, A\&A, 488, 401

\bibitem[Ruchti et al.(2013)]{Ruchti}
Ruchti G.R., Bergemann M., Serenelli A., Casagrande L., Lind K., 2013, MNRAS,
429, 126

\bibitem[Salaris(2013)]{Salaris}
Salaris M., 2013, in {\it Advancing the physics of cosmic distances}, IAU
Symposium 289, R. de Grijs ed, Cambridge University Press

\bibitem[Schlegel et al.(1998)]{Schlegel}
Schlegel D.J., Finkbeiner D.P. \& Davis M., 1998, ApJ, 500, 525

\bibitem[Sch\"onrich et al.(2012)]{SBA}
Sch\"onrich R., Binney J., Asplund M., 2012, MNRAS, 420, 1281 (SBA)

\bibitem[Sch\"onrich et al.(2012)]{SchoenrichBD}
Sch\"onrich R., Binney J., Dehnen W., 2012, MNRAS

\bibitem[Sharma et al.(2011)]{Sharma11}
Sharma S., Bland-Hawthorn J., Johnston K.V. \& Binney J., 2011, ApJ, 730, 3

\bibitem[Siebert et al.(2011)]{DR3}
Siebert A., Williams M.E.K., \& the RAVE collaboration, 2011, AJ, 141, 187

\bibitem[Steinmetz(2006)]{RAVE} Steinmetz, M. et al., 2006, AJ, 132, 1645

\bibitem[Strutskie et al.(2006)]{2MASS} Strutskie M.F., et al., 2006, AJ,
131, 1163

\bibitem[van Leeuwen(2007)]{vanLeeuwen}
van Leeuwen F.,  2007,
	\textit{Hipparcos, the New Reduction of the Raw Data}, Springer Dordrecht

\bibitem[Williams et al.(2013)]{Williams13}
Williams M.E.K., et al., 2013, MNRAS, in press (arXiv1302.2468)

\bibitem[Yanny et al.(2009)]{SEGUE}
Yanny B. et al., 2009, AJ, 137, 4377

\bibitem[York et al.(2000)]{SDSS}
York D.G., et al., 2000, AJ, 120, 1579

\bibitem[Zwitter et al.(2008)]{DR2}
Zwitter T., et al., 2008, AJ, 136, 421

\bibitem[Zwitter et al.(2010)]{Zwitter10}
Zwitter T., et al., 2010, A\&A, 522, 54
\end{thebibliography}
\end{document}